\documentclass[12pt, lettersection]{article}
\usepackage[dvips]{graphicx}
\usepackage{amssymb,amsfonts,amsmath}
\usepackage{graphicx}
\usepackage{dcolumn}
\usepackage{epstopdf}
\usepackage{bm}
\usepackage{graphicx}
\usepackage{subfigure}
\usepackage{color}
\usepackage{xcolor}
\usepackage{latexsym}
\usepackage{amsfonts}
\usepackage{amssymb,txfonts, pxfonts}
\usepackage{amssymb,amsfonts,amsmath}

\usepackage{setspace}
\usepackage[top=25mm, bottom=25mm, left=25mm, right=25mm]{geometry}

\newcommand \beq{\begin{equation}}
\newcommand \eeq{\end{equation}}
\newcommand \beqa {\begin{eqnarray} } 
\newcommand \eeqa {\end{eqnarray}}

\newcommand \half {\frac{1}{2}}
\newcommand \Ocal  {\mathcal{O}}

\newcommand \deff {\equiv}
\newcommand \dd {\partial}
\newcommand \h {\half}
\newcommand \goes {\approx}
\newcommand \like {\sim}
\newcommand \ra {\rangle}
\newcommand \la {\langle}

\newcommand \xtwo {\la x^2 \ra}
\newcommand \xfour {\la x^4 \ra}
\newcommand \xthree {\la x^3 \ra}
\newcommand \xone {\la x \ra}

\newcommand \xbar {\xone}
\newcommand \xbart {\la x(t) \ra}
\newcommand \xbartmin {\la x(t_{min}) \ra}

\newcommand \ib {{I_B}}

\newcommand \bd {{B_D}}
\newcommand \br {{B_R}}

\usepackage[numbers]{natbib}

\begin{document}

\title{Response to a population bottleneck can be used to infer recessive selection}

\author{
Daniel J.~Balick\thanks{Brigham and Women's Hospital, Department of Medicine, Harvard Medical School} \thanks{Broad Institute of Harvard and MIT}, 
\  Ron Do{$^\dag$} \thanks{Department of Genetics, Harvard Medical School} \ \thanks{Massachusetts General Hospital, Department of Medicine, Harvard Medical School} , 
\ David Reich{$^{\dag \ddag}$}, 
\ and Shamil R. Sunyaev{$^{*\dag}$}
}
\date{}



\maketitle

\begin{abstract}
Here we present the first genome wide statistical test for recessive selection.  This test uses explicitly non-equilibrium demographic differences between populations to infer the mode of selection.  By analyzing the transient response to a population bottleneck and subsequent re-expansion, we qualitatively distinguish between alleles under additive and recessive selection.  We analyze the response of the average number of deleterious mutations per haploid individual and describe time dependence of this quantity.   We introduce a statistic, $\br$, to compare the number of mutations in different populations and detail its functional dependence on the strength of selection and the intensity of the population bottleneck.  This test can be used to detect the predominant mode of selection on the genome wide or regional level, as well as among a sufficiently large set of medically or functionally relevant alleles.
\end{abstract}


%

\newpage

\section{Introduction}

In diploid organisms, selection on an allele, or a group of alleles, can be categorized as additive, dominant or recessive, or as part of a more general epistatic network. A large body of existing work is devoted to statistical methods to detect and quantify selection using DNA sequencing data, including comparative genomics and the sequencing of population samples \cite{Cutter2013,Zhang2012,Hill2008}. However, much less progress has been made toward identifying the predominant mode of selection as additive, recessive or dominant. Genetics of model organisms and of human disease provide plenty of anecdotal evidence in favor of the importance of dominance \cite{Lynch1998}. Although genome-wide association studies suggest that alleles of small effects involved in human complex traits frequently act additively, estimation of genetic variance components from large pedigrees suggests a substantial role for dominance in a number of human quantitative traits \cite{Newman2001}. Alleles of large effects involved in human Mendelian diseases, spontaneous and induced mutations in model organisms, such as mouse, zebrafish, or Drosophila, are frequently recessive \cite{Herron2002}. In spite of these observations, the role of dominance in population genetic variation and evolution remains unexplored and no formal statistical framework to test for dominance coefficient is currently available.  

Using a combination of theoretical analysis and computer simulations, we demonstrate that recessive selection can be qualitatively distinguished from additive selection in populations that experienced a population bottleneck and subsequent re-expansion. 
Previous studies of non-additive variation in the presence of a bottleneck lack a complete description of the dynamics after re-expansion \cite{Robertson1952,Bryant1986,Wang1998,Zhang2004,Hill2008}, or focus on epistatic interactions rather than recessive selection \cite{Cheverud1996, Hill1998, Naciri2003, Barton2004, Hill2006, Turelli2006}, with the notable exception of a recent independently conducted complementary analysis found in \cite{Simons2013}.
Contrary to na\"{i}ve expectation, the number of deleterious recessive alleles per haploid genome is transiently reduced after a population bottleneck, while the number of additively or dominantly acting alleles is increased. In spite of a well-documented increase in frequency of some recessively acting variants in founder populations, the average number of recessive alleles carried by an individual is reduced as a consequence of the bottleneck. With the growing availability of DNA sequencing data in multiple populations, these results demonstrate the potential to directly evaluate the role of dominance, either on a whole genome level, or in specific categories of genes. 

\begin{figure}
\begin{center}
\includegraphics[width=0.65\columnwidth]{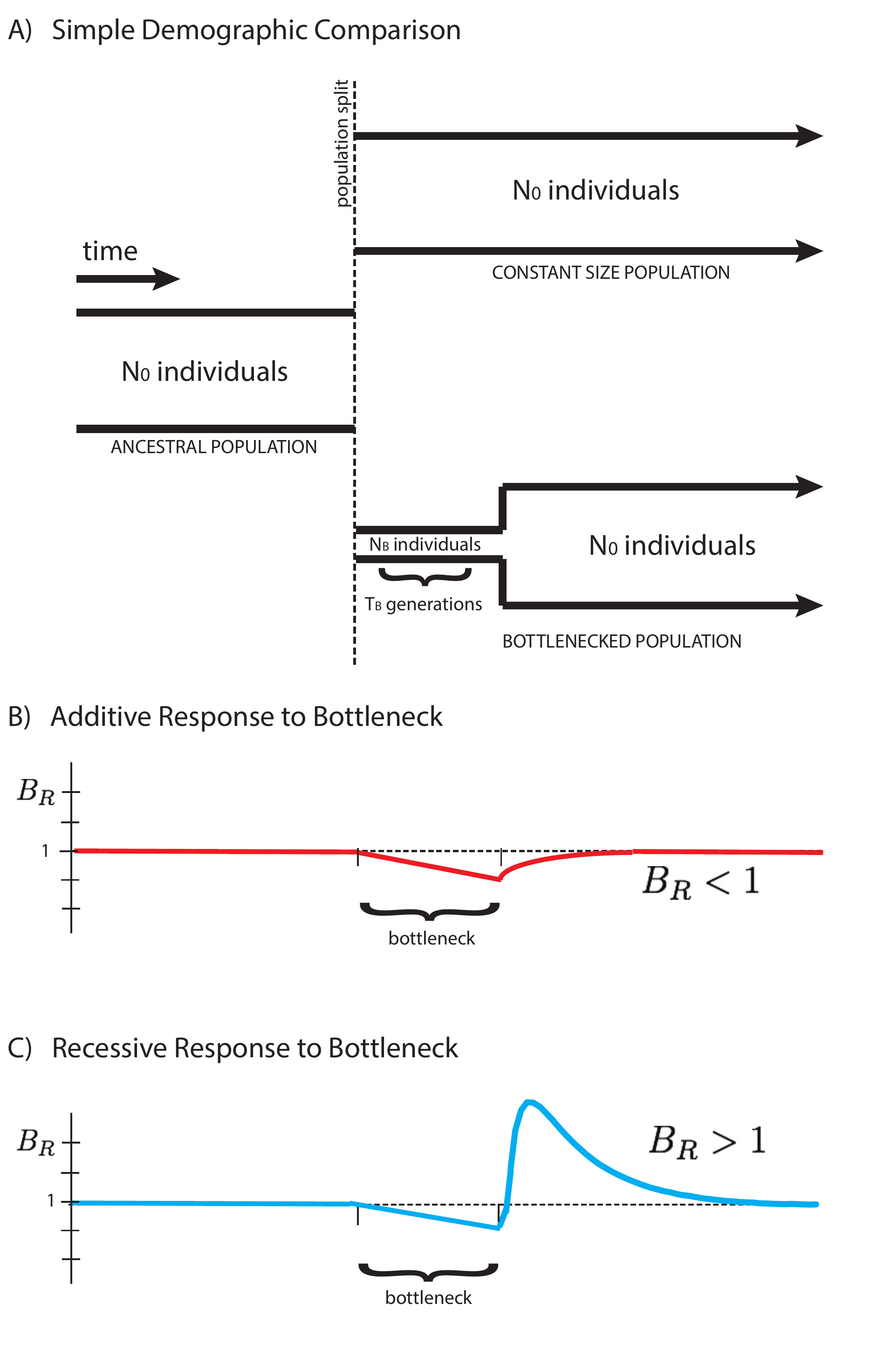}
\end{center}
\caption{{A schematic representation of two populations is presented above ({\bf A}).  Initially a single population prior to the bottleneck event, the populations split and have distinct demographic profiles.  The equilibrium population has constant size for easy comparison to the founded population.  The latter drastically reduces its population size to $N_B$ for a short time $T_B$ during the founder's event.    Our statistical comparison between populations $\br$ is represented here for cases of purely additive ({\bf B}) and purely recessive ({\bf C}) variation.  The statistic $\br>1$ for recessive variation (dominance coefficient $h=0$) and $\br<1$ for additive variation ($h=1/2$), providing a simple test for the primary mode of selection of polymorphic alleles in the populations.}\label{fig:bottle0}}
\end{figure}

Population bottlenecks are a common feature in the history of many human populations. For example, the ``Out of Africa" bottleneck involved ancestors of many present-day human populations. Numerous recent bottlenecks affected, among others,well studied populations of Finland and Iceland. More generally, bottlenecks followed by expansions are standard features in the recent evolution of most domesticated organisms. We suggest that complex demographic history may assist rather than complicate statistical inference of selection in population genetics.  Here we use the distinct demographic histories of two subpopulations to identify the type of selection dominating the dynamics, and show that the average number of mutations per individual, $\xbar$, is dependent on the mode of selection.  We introduce a measure $\br$ (the ``burden ratio" defined below) that provides a simple statistical test for any set of polymorphic alleles in the population, where $\br<1$ corresponds to predominantly additive selection and $\br > 1$ to predominantly recessive selection, as shown in Figure \ref{fig:bottle0}.  This test is not restricted to the simplified demographic model presented in this paper, but rather provides a quite generic qualitative test for the predominance of recessive selection in comparison between two populations, one of which experienced a bottleneck event.
\newline

\section{Model}
We work with a simple demography described by an ancestral population of $N_0$ individuals that splits into two subpopulations, one with population size $N_0$ equal to the initial population size (``equilibrium"), and one with reduced bottleneck population size $N_{B}$ (``founded").  The latter population persists at this size for $T_B$ generations before instantaneously re-expanding to the initial population size $N_0$, as shown in Figure \ref{fig:bottle0}.  Time $t$ is measured after the re-expansion from the bottleneck, as we are interested in the dynamics during this period.  Quantities measured in the equilibrium population, and equivalently prior to the split, are denoted with a subscript ``$_0$".  We consider only deleterious mutations with average selective effect of magnitude $s > 0$, such that $s$ represents the strength of deleterious selection. Extensions of this analysis to a full distribution of selective effects can be found in the {\bf SI}. The initial population is in steady state with $2N_0 U_{d}$ deleterious alleles introduced into the population at a mutation rate $U_{d}$ per haploid individual per generation.  In a steady state equilibrium, the site frequency spectrum (SFS) of polymorphic alleles is given by Kimura \cite{Kimura1964}.
\beq
\label{kimura}
\phi_{eq}(x) =  4  N  U_{d} \frac{e^{-4 N s h x-2 N s (1-2h) x^2}}{x \left(1-x \right)}   \left[ 1 -   \frac{  \int^x_0 dy \ e^{4 N s h y+2 N s (1-2h) y^2} }{  \int^1_0 dy \  e^{4 N s h y+2 N s (1-2h) y^2} }\right]
\eeq
Here $h > 0$ is the dominance coefficient for deleterious mutations, where $h = 1/2$ corresponds to a purely additive set of alleles, and $h=0$ corresponds to the purely recessive case.  For the present analysis, we primarily focus on these two limits, contrasting their effects on the genetic diversity.  The solution represents a mutation-selection-drift balance in which new mutations are exactly compensated for by the purging of currently polymorphic alleles due to selection and extinction due to stochastic drift.    In this way, an approximately static number of polymorphic alleles exists in the population at any given time.  

\section{Results}
We follow the expected number of mutations per chromosome in the population, which is simply the first moment of the SFS.
\beq
\xbar = \int  x \  \phi(x) 
\eeq
When multiplied by $s$, this is the effective ``mutation load" of each individual in the additive case, but in the case of purely recessive selection this is not proportional to the fitness, as selection acts only on homozygotes.  We refer to this statistic generally as the ``mutation burden" to avoid assumption of any given mode of selection.  Comparison between the mutation burden in the equilibrium and founded populations in the form of the ``burden ratio", $\br$, provides a test for recessive alleles.
\beqa
\br \deff \frac{\xbar_{eq}}{\xbar_{founded}} = \left\{
        \begin{array}{ll}
            < 1 & \text{for additive selection} \\ \\
             > 1  & \text{for recessive selection}
        \end{array}
\right.
\eeqa
To gain intuition for this qualitative difference,  we work to quantitatively understand the population dynamics in a simple demography, first for purely additive selection, and then for purely recessive selection for comparison.

\subsection{Additive selection and response to a bottleneck}
The initial site frequency spectrum $\phi^{A}_0(x)$ for purely additive alleles is given by Equation (\ref{kimura}) with $h = 1/2$.  
\beq
\label{kimura_add}
\phi^{A}_0(x) =  \frac{\theta_0 }{x(1-x)} \frac{1-e^{ 2 N_0 s (1-x)}}{1-e^{2 N_0 s}} 
\eeq
Here $\theta_0 = 4 N_0 U_{d}$.  When $2 N_0 s \gg 1$, the SFS rapidly decays as $x\rightarrow 1$ simplifying the functional form\cite{Nei1968}.  We approximately compute the initial mutation burden as follows.
\beq
\xbar_0 \approx \theta_0  \int_0^1 x  \ \frac{e^{-2 N_0 s x}}{x } \approx \frac{2U_{d}}{s} 
\eeq
Now we deviate from equilibrium by reducing the population size to $2N_B$ chromosomes, representing a population bottleneck.  The effect that a bottleneck has on the site frequency spectrum is twofold: a fraction of alleles are removed from the population due to increased random drift, and the mean of the remaining alleles occurs at higher frequency. 
The dynamics of the distribution $\phi(x,t)$ during such a change in demography can be computed from Kolmogorov's forward equation, as detailed in the {\bf SI}. 
The first moment of the distribution, the mutation burden, follows the temporal dynamics derived from summing the Kolmogorov equation over all alleles in the genome, and takes the following form.
\beq
\label{add_det_0}
\dd_t \xbar \approx U_{d} - \frac{s}{2} \left(\xbar- \xtwo \right) 
\eeq
The burden of additive mutations is not directly affected by drift, as the drift term vanishes from the dynamics of the first moment, however the dependence on the second moment introduces an indirect dependence on drift. 
In the strong selection regime, in the limit where $\xtwo \ll \xone$, extinction of some alleles is exactly compensated for by an increase in frequency of other alleles. This is true in the equilibrium distribution prior to the bottleneck when $N_0 s \gg 1$, where $\xbar_0 \sim \Ocal \left(U_d/s\right)$ and $\la x^2_0 \ra \sim \Ocal \left(U_d/(N_0s^2)\right)$.  During the bottleneck, the mutation burden $\xbar$ monotonically increases; the second moment $\xtwo$ increases, as well, reaching a maximum value in the case of a long bottleneck where it scales as $\xtwo \sim \Ocal (U_d/(N_Bs^2))$.  Provided $N_B s\gg 1$, the second moment is guaranteed to be subdominant to the first moment, simplifying the dynamics as follows.
%
\beq
\label{add_det}
\dd_t \xbar \approx U_{d} - \frac{s}{2} \xbar 
\eeq
For a bottleneck of duration $T_B$, this equation admits solutions of the form,
\beqa
\nonumber \\
\la x(T_B) \ra \approx \xbar_0 \ e^{- \frac{sT_B}{2} } + \frac{2U_{d}}{s}(1- e^{- \frac{sT_B}{2}}) = \frac{2U_{d}}{s}.
\eeqa
After plugging in the initial value $\la x(0) \ra = \xone_0 = 2U_{d}/s$, we find that the time dependence drops out completely, demonstrating that the population remains in mutations selection balance throughout the bottleneck.  After instantaneous re-expansion to the initial population size, the dynamics of the distribution $\phi(x)$ are completely analogous to those inside the bottleneck in this limit, such that the mutation burden never deviates during the demographic perturbation.  
%


In the opposite limit of completely relaxed selection during the bottleneck, the dynamics of the mutation burden are completely driven by the influx of new mutations.
\beqa
\dd_t \xone= U_{d}
\eeqa
The net effect of this accumulation over the course of the bottleneck is simply the integral of this quantity.  For a bottleneck with duration $T_B$ generations, the net effect of mutation accumulation due to relaxed selection is given simply by the following expression.
\beqa
 \la x (T_B) \ra \approx \xone_0 + U_{d} T_B
\eeqa
Additionally, one can show that the second non-central moment gains an analogous contribution in addition to the net effect of drift.
\beqa
 \la x^2 (T_B) \ra \approx \xtwo_0 + I_B \xone_0 + U_{d} I_B
\eeqa
Here we have expressed the second moment as a function of the bottleneck intensity $I_B \deff \frac{T_B}{2N_B}$.  Immediately after re-expansion from the bottleneck, selection is again efficient, such that the dynamics are completely described by Equation (\ref{add_det_0}).  Although the second moment is increased due to relaxed selection during the bottleneck, we find that this increase is negligible in comparison to the direct accumulation in the first moment provided $I_B \ll 1$.  As a result, the primary effect of the bottleneck in this limit is to accrue new mutations that are subsequently purged when selection is again efficient in the re-expanded population.  The dynamics for the two limiting cases can be summarized as follows.
\beqa
\xbart_{founded} \approx  
\left\{
        \begin{array}{ll}
           \frac{2 U_{d}}{s} & \quad  \text{for} \ \ \ 2 N_B s \gg 1 \\ \nonumber \\
            \frac{2 U_{d}}{s} + U_{d} T_B e^{- \frac{st}{2} }  & \quad   \text{for} \  \ \  2N_B s \ll1, \  \ \ I_B \ll 1
        \end{array}
    \right.
\eeqa
We note that $\xbar_{founded} \ge \xbar_{eq}$ at all times in both limiting cases, and asymptotically decays to the equilibrium frequency on a timescale given by the strength of selection of the accumulated deleterious mutations. In the case of an instantaneous bottleneck, we find that the mutation burden is only slightly shifted even if selection is fully relaxed, resulting in effectively no observable change in either limit.  Our statistical measure, the burden ratio $\br$, in the additive case can be written approximately as follows.
\beqa
\br^{additive}(t) =  \frac{\xbar_{eq}}{  \xbar_{founded}} \approx  
\left\{
        \begin{array}{ll}
          1 & \quad  \text{for} \ \ \ 2 N_B s \gg 1 \\ \nonumber \\
            \left(1+ \frac{s T_B}{2}  e^{- \frac{st}{2} } \right)^{-1} \ge 1 & \quad   \text{for} \  \ \  2N_B s \ll1, \  \ \ I_B \ll 1
        \end{array}
    \right.
\eeqa
As we will see in the following sections, recessive selection results in depleted mutation burden with corresponding values $\br > 1$, proving a contrast to the additive scenario and justifying our use of this statistic as a test for recessivity.

\subsection{Recessive selection and dynamics of the mutation burden}

Prior to the bottleneck, the initial site frequency spectrum for alleles under recessive selection is given by the $h=0$ limit of Equation (\ref{kimura}).  
\beq
\label{kimura_rec}
\phi^{R}_0(x) = \theta_0 \frac{e^{-2 N_0 s x^2}}{x \left(1-x \right)}  \left[ 1 -   \frac{  \int^x_0 dy \ e^{2 N_0 s  y^2} }{  \int^1_0 dy \ e^{2 N_0 s  y^2} }\right]
\eeq
At low frequencies $x < \sqrt{4 N_0 s}$ the spectrum decays slower than in the additive case, representing alleles protected from recessive selection by existing primarily in heterozygous form.  In contrast, at high frequencies the spectrum decays faster than the additive exponential decay, falling off as $e^{-2N_0 s x^2}$.

\subsubsection{Instantaneous population bottlenecks}
First, we restrict our analysis to an instantaneous bottleneck with intensity $I_B = 1/2N_B$, as this provides insight into the non-equilibrium response of the frequency spectrum to a downsampling event.  Later, we extend our analysis to finite bottlenecks that persist for $T_B$ generations, with total intensity $I_B = T_B/2N_B$.
We represent the increase in drift due to a single generation bottleneck by downsampling.  During this time step, $N_B$ diploid individuals are chosen at random from the initial larger population of $N_0$ individuals.  
\beq
\label{phib}
\phi_B(k, t_B = 0) = \binom{2 N_B}{k} \int dy \   (1-y)^{2 N_B - k} (y)^k \phi_0(y)
\eeq
Binomial sampling gives the distribution $\phi_B$ of deleterious alleles with frequency $x=k/2N_B$. There is a loss of allelic variation due to the bottleneck, corresponding to the $k=0$ term in Equation (\ref{phib}).

Re-expansion is modeled as up-sampling the distribution $\phi_B(x)$ from $N_B$ to $N_0$ diploid individuals, which has negligible effect on the first and second moments of the distribution.   As a result of drift to higher frequencies during the bottleneck, much of the existing variation appears in homozygous form immediately after the increase in population size. These individuals are rapidly selected out of the population, driving high frequency alleles to lower frequencies on a very short time scale.   Since drift is once again suppressed, selection becomes far more efficient, particularly for alleles of large selective effect. 

The time evolution of $\phi$ after the bottleneck is given by the forward Kolmogorov equation for recessive selection (see {\bf SI}).
The mutation burden follows the time dependence,
\beq
\label{mean_motion}
\dd_t \langle x(t)\rangle \approx  U_{d} - s \langle x(t)^2 \rangle.
\eeq
Here we suppress a selection term proportional to $\xthree$ of $\Ocal(1/\sqrt{Ns})$ in analogy to the additive case.
Since recessive selection depends quadratically, rather than linearly, on the allele frequency, the increased variance of the distribution drives the motion of the mutation burden.    Alleles with frequency $x > \sqrt{1/2N_0}$ appear in homozygous form and are rapidly pushed down to lower frequencies.  This happens on a time scale of order $s^{-1/2}$ and effectively reduces the variance, slowing the decrease in the mutation burden $\xbar$.  New mutations introduced during this period slowly drift to appreciable frequencies, replacing those lost in the bottleneck.  This process is drift controlled, rather than selection controlled, and thus occurs on a time scale of $\Ocal(2N_0)$ generations.  As a result, the mutation burden quickly decreases due to selection immediately after the bottleneck until it slows to a stop, and then gradually increases as the population accumulates new mutations and re-equilibrates.

A minimum in the mutation burden $\xbart_{founded}$ occurs when the time derivative vanishes.  This corresponds to a characteristic time scale associated with the selective effect $s$, where our statistical test $\br = \frac{\xbar_{eq}}{ \xbar_{founded}}$ is maximized.  Since this time scale is shorter than the time scale of drift, we can imagine rescaling time by the effective population size $2N_0$ and then working in the perturbative regime $t/2N_0 \ll 1$.  This allows us to Taylor expand near the re-expansion time $t = 0$ to understand the motion of the mutation burden at times soon after the bottleneck.  
\beq
\label{taylor}
\dd_t \la x (t) \ra \goes U_{d} - s \left[ \langle x(t)^2 \rangle |_{t=0} + t \dd_t \langle x(t)^2 \rangle |_{t=0} + \frac{t^2}{2} \dd_t^2 \langle x(t)^2 \rangle  |_{t=0} + \Ocal(t^3) \right]
\eeq
To understand the time dependence of $\la x^2 \ra$, specifically the time derivative, we analyze the higher moments in the same fashion as employed for the first moment in Equation (\ref{mean_motion}).  All relevant moments are computed in the {\bf SI} and we note sufficient convergence to validate this expansion. This allows for the re-expression of Equation (\ref{taylor}) to second order in $t$ in terms of the first three moments of the site frequency spectrum immediately after re-expansion.  The moments of the post-bottleneck initial distribution can be written in terms of the initial equilibrium distribution using the integral form given in Equation (\ref{phib}).   Details of this calculation appear in the {\bf SI}.  In the strong selection limit $2N_0 s \gg 1$ these initial equilibrium moments are readily approximated by standard convolutions of a polynomial with a Gaussian.  Suppressing subdominant contributions in the limit $N_B^2 \gg N_0 s$, we find the following approximation to the trajectory of the mutation burden immediately after the bottleneck re-expands.   
\beq
\xbart \like U_{d}\frac{ \sqrt{4 N_0}}{\sqrt s}\left( 1- \frac{s t }{2 N_B} \right) +  U_{d}\frac{3 s t^2}{2N_B} + \Ocal(t^3)
\eeq
Concentrating on this second order expansion in $t$, we find that the curve first drops from its initial value $\la x (0) \ra =U_{d} \sqrt{ \frac{4 N_0}{ s}}$, quickly reaches a minimum, and is then brought back up by the the positive second order term.  The location of the minimum is easily found to have the following parameter dependence.
\beq
\label{tmin}
t_{min} \propto \sqrt{\frac{4N_0}{s}}
\eeq
The second derivative is positive at this extremum, implying a local minimum.  Plugging $t_{min}$ into our expression for $\xbart$ in the limit $N_0 s \gg 1$, we find the following minimum value for the average number of recessive deleterious mutations per genome following a bottleneck.
\beq
\xbartmin \sim  \theta_0 \left( \frac{1}{\sqrt{4N_0 s}}  - \frac{1}{24 N_B} \right)
\eeq
We note that $\xbar_0 \like \frac{\theta_0}{\sqrt{4N_0 s}}$ is the approximate mutation burden for the equilibrium distribution in the $2N_0 s \gg 1$ limit, allowing us to simply write the extreme value of the $\br$ statistic as follows.  
\beq
\br(t_{min}) \sim   \left(1-\frac{\sqrt{ 4 N_0 s}}{24N_B}\right)^{-1} > 1
\eeq
We find the following dependence on time in immediate response to a population bottleneck.
\beq
\label{Dtime}
\br^{recessive} (t)\sim  \left(1- \frac{s t}{2 N_B}  + \frac{3 s^{3/2}  t^2 }{2N_B \sqrt{4N_0}} + \Ocal(t^3) \right) ^{-1} > 1
\eeq
This expansion is only valid in the small time limit where the quadratic term is subdominant, such that all values are positive.  Long before this simple quadratic expression becomes negative, higher order contributions become relevant and dominate.  As seen in simulations described in the following section, for recessive deleterious mutations, the burden ratio remains positive at all times.

This precise result applies strictly in the limit of a strong, single generation bottleneck, where $N_0 \gg N_B$.  Additionally, the technique used to compute integral expressions required the strong selection limit $2 N_0 s \gg1$.   Analysis of higher order contributions to the trajectory are made substantially easier by restricting to the limit $2N_B > \sqrt{2 N_0 s}$, which happens to be biologically reasonable, for example, in human populations where most examples of founding events are on the order of $N_0 \sim 10^4$ and $N_B \sim 10^3$ (see further discussion in the $SI$ on general dominance coefficients).  Despite these analytic restrictions in parameter space, our simulations described below indicate that the signature of $\br > 1$ is ubiquitous for populations under predominantly recessive selection.

\subsubsection{Extended population bottlenecks}
We argue that for the case of relatively low intensity bottlenecks, where intensity is defined as $I_B \deff T_B/2N_B \ll 1$, we can approximately express the magnitude of $\br$ using a simple substitution $(2N_B)^{-1} \rightarrow I_B$.  This is equivalent to the claim that for low intensity bottlenecks, the $\br$ statistic depends only on the ratio of the bottleneck time to the bottleneck population size, and any explicit dependence on $T_B$ occurs in subdominant contributions.  This intuition is confirmed by simulations described in below, where we show that the accuracy of our analytic approximation breaks down as $I_B \rightarrow 1$ and the intensity becomes non-perturbative.  For short bottlenecks with $I_B < 1/10$, the approximation of an instantaneous single generation sampling event remains sufficiently accurate, even for strong selective coefficients $s\sim0.1$.  Under this trivially extended instantaneous approximation, $\br(t)$ can be written in terms of the intensity of a short bottleneck as follows.
 \beq
 \label{IB}
\br^{extended} (t)\sim  \left(1- I_B \left(s t -\frac{3  s^{3/2}  t^2 }{ \sqrt{4N_0}}  + \Ocal(t^3) \right)\right) ^{-1} > 1
\eeq
The $\br$ of maximum effect, has a magnitude given approximately by,
\beq
\label{IBmax}
\br^{extended}(t_{min}) \sim    \left(1-I_B \frac{\sqrt{ N_0 s}}{6 }\right)^{-1} .
\eeq

\begin{figure}
\begin{center}
\includegraphics[width=0.7\columnwidth]{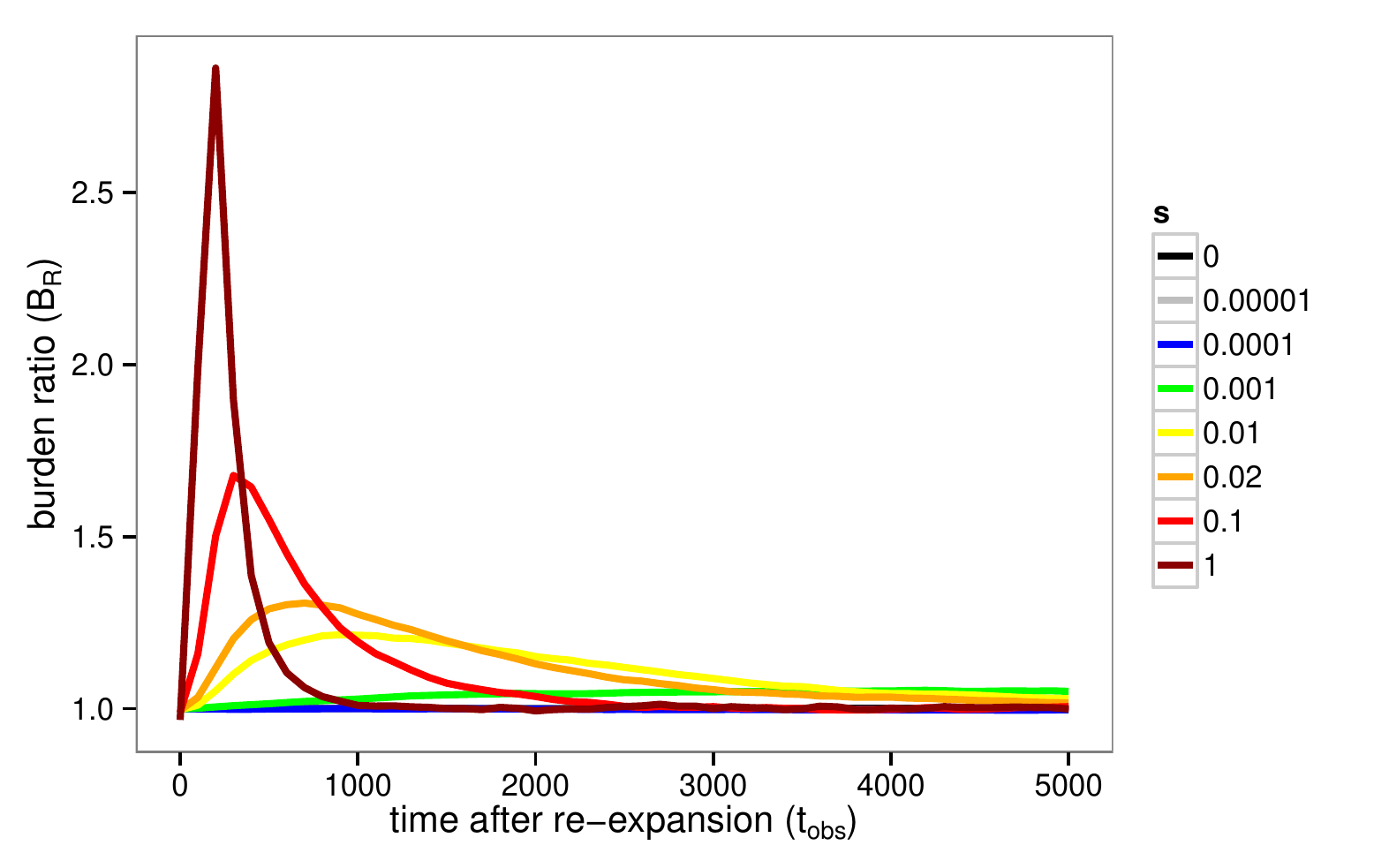}
\end{center}
\caption{{The time dependence of $\br(t)$ after a population bottleneck is shown for various selective coefficients.  Peak $\br$  values vary in both magnitude and time as a function of $s$.  The founded population was simulated with $2N_0 = 20000$, $2 N_B = 2000$, and $T_B = 200$ and plotted for $5000$ generations after re-expansion.}\label{fig:Dtime}}
\end{figure}

For illustration of the behavior described in the above analytics we present a time series of recessive simulations with curves representing various selection coefficients in Figure \ref{fig:Dtime}.  The time dependence of the $\br$ statistic is plotted to demonstrate the simulated population's response to a founder's event.  Crucially, we find that the peak $\br$ values vary in both magnitude and time as a function of $s$, as is consistent with our analytic understanding and intuition.  

\subsection{Transient response and time of observation determine detectable selection coefficients}
\label{subsection:transient}
Thus far, we have detailed the dynamic dependence of a set of alleles in a population, all with selective effect $s$, in response to demographic perturbation in the form of a bottleneck.  Notably, for recessive selection, a peak response occurs in the $B_R$ statistic at some time $t_{min}$ after re-expansion.  In general, both the magnitude of $B_R(t_{min})$ and the time of the peak itself depend sensitively on the selection coefficient.  In general, a distribution of mutations with different selective effects will be present, many of which may be simultaneously polymorphic in a given population.  Since alleles of different selective effect respond to the bottleneck on different time scales, one can ask what selective effect is most likely to be observed at a given time.  For example, very strong selection has the tendency to peak and subsequently re-equilibrate immediately after the bottleneck, such that observation of alleles with large $s$ is substantially more difficult at later times.  On the other hand, alleles under relatively weak selection have a peak effect at very late times, such that at the time of data collection a statistically significant response may not yet have occurred.   

\begin{figure}
\begin{center}
\includegraphics[width=0.7\columnwidth]{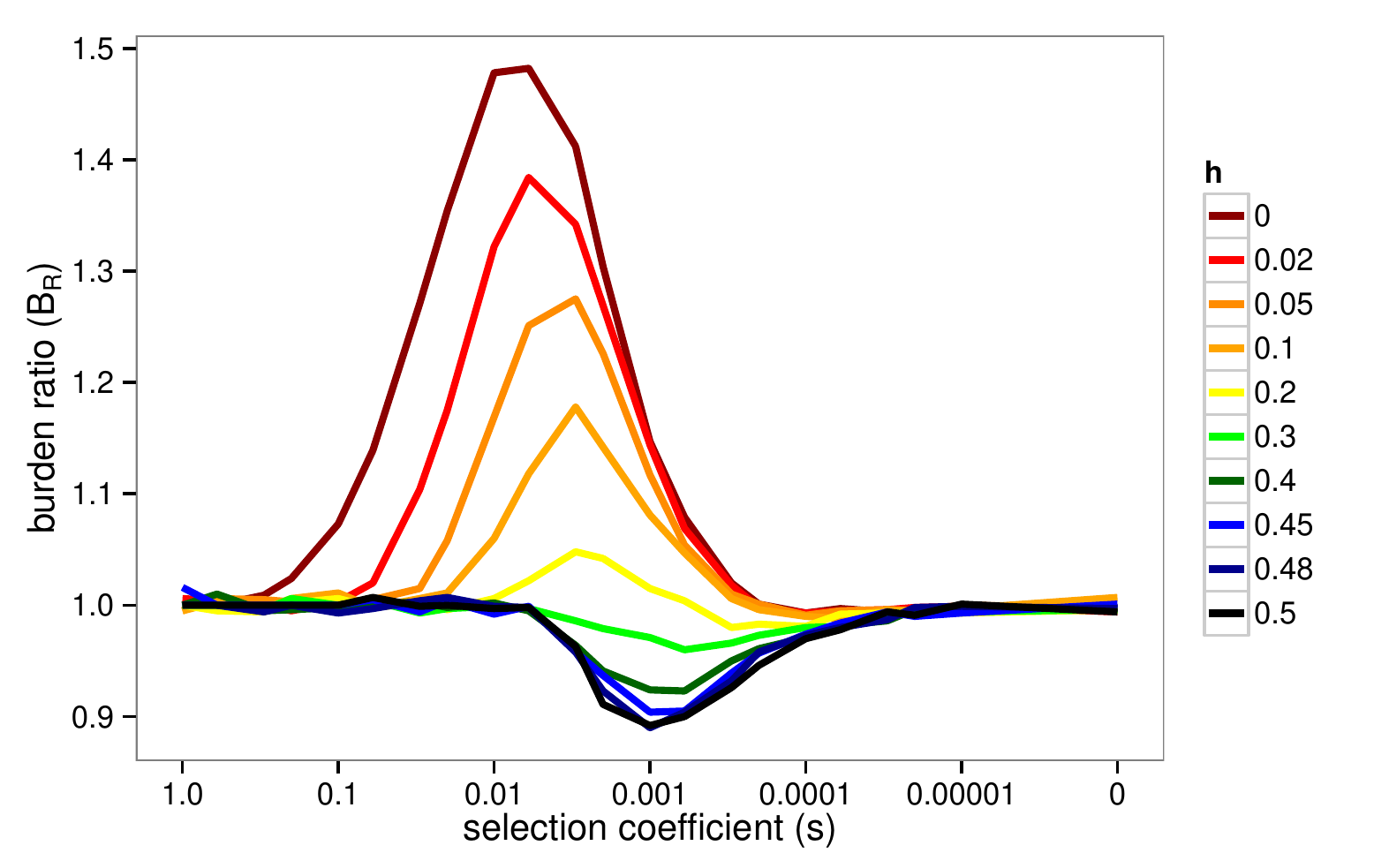}
\end{center}
\caption{{At the time of observation $t_{obs}$, the value of $\br (t_{obs})$ is determined by the average strength of selection $s$ for additive or recessive variation, or variation with any intermediate dominance coefficient $h \in [0,\h]$.  A range of $\br$ values observed at a single time slice are plotted for various $s$ values.  Different dominance coefficients appear as solid lines with fully recessive selection ($h=0$) at the top and purely additive selection ($h=\frac{1}{2}$) at the bottom.  $\br$ approaches one both in the limit of very strong selection $s\rightarrow1$ due to the rapid transient response, and in the very weak selection limit $s\rightarrow 0$ due to the nearly neutral insensitivity to the bottleneck.  For some intermediate dominance coefficients $h_c$, a crossover occurs ($h_c\sim \frac{1}{4}$ in the example shown) where the effects of additive and recessive variation cancel such that $\br(h_c) \sim1$.   The parameter dependence of the crossover is explored analytically in the {\bf SI}.}
\label{fig:t_obs}}
\end{figure}

We would like to understand the transient behavior of the burden ratio $\br(t)$, as well as the value of the selection coefficient $s$ for which $\br$ is largest at a given time.  When comparing to population data, one has little control over the demographic history, and thus it becomes important to understand the selective coefficient that dominates at the time of observation.  According to the time dependent expression in Equation (\ref{IB}), we expect the effect to decrease quite rapidly for very large $s$.  However, the peak occurs quite early in the case of larger $s$ values, allowing the mutation burden to equilibrate over a longer period of time between the peak and observation to return to mutation burden values close to $\br\like 1$.  This tells us that the equilibration process is what reduces the magnitude of $\br $ for large $s$.   In the case of very recent bottlenecks, the large $s$ values dominate, but for later times of observation, this signal has partially equilibrating, potentially allowing a smaller $s$ value to dominate the statistic.  
At a given time of observation $t_{obs}$, one can represent $\br(s,t_{obs})$ as a function of various selection coefficients $s$.  Figure \ref{fig:t_obs} represents $\br(s)$ for a fixed $t_{obs}$ for various dominance coefficients $h$.  We concentrate here on recessive variation with $h=0$, but note that a crossover occurs at some value $h_c$ where additive and recessive effects offset each other in the $\br$ statistic (detailed in {\bf SI}). 
Based on our analytics, we expect the peak to shift from extreme high $s$ values at early times to extreme low $s$ values at late times, eventually dissolving into neutrality.  
We take the $s$ derivative of Equation (\ref{IB}) to find the maximum at $t_{obs}$.
\beq
\dd_s \br(s,t_{obs})  |_{s = s_{max}} \propto  -\ib  t_{obs}  + \frac{9 s^{1/2} \ib t_{obs}^2 }{2 \sqrt{4N_0}} =0
\eeq
\beq
\label{smax}
s_{max} \sim \frac{16N_0}{81t_{obs}^2} \sim \frac{2N_0}{10t_{obs}^2}
\eeq
One can easily show that the second derivative evaluated at this point is negative, confirming that this is a maximum.  This result matches our intuition: maximum $s$ values of $\br(s,t)$ are found at high $s$ for early times,  $s_{max}(t\rightarrow 0) \gg 1$, and at low $s$ for late times, $s_{max}(t\rightarrow \infty) \ll 1$.  This is qualitatively observed in our simulations by comparing the relative values of $\br(s)$ as a function of time.  

As the effect is transient, we can define a relaxation time $t_{relax}$ corresponding to the vanishing of any response to the bottleneck.  This is given by determining when $s_{max}$ is dominated by effectively neutral variation at roughly $s_{max}\sim 1/2N_0$.  After this time, $\br(s,t)$ cannot be differentiated from one for any $s$. 
\beq
t_{relax} < \frac{2N_0}{\sqrt{10}}  < 2N_0
\eeq
We note that the return to equilibrium happens on a time scale faster than random drift, even for the weakest selective effects, thus validating our perturbative approximations using $t/2N_0 \ll1$. Higher order time dependence in Equation (\ref{IB}) may substantially correct this estimate, but we feel that the presentation of this methodology is conceptually important and provides a greater understanding of the transient dynamics of population response to bottlenecks.  As it is relevant to human populations, we note that if both populations expand exponentially after the bottleneck, the effect may persist long beyond $t_{relax}$.  This is explored analytically in the {\bf SI} and in simulations in an accompanying paper \cite{Do2014}.


\section{Comparison of analytic results to simulation}
\label{sec:sim_2}
\begin{figure}
\begin{center}
\includegraphics[width=0.8\columnwidth]{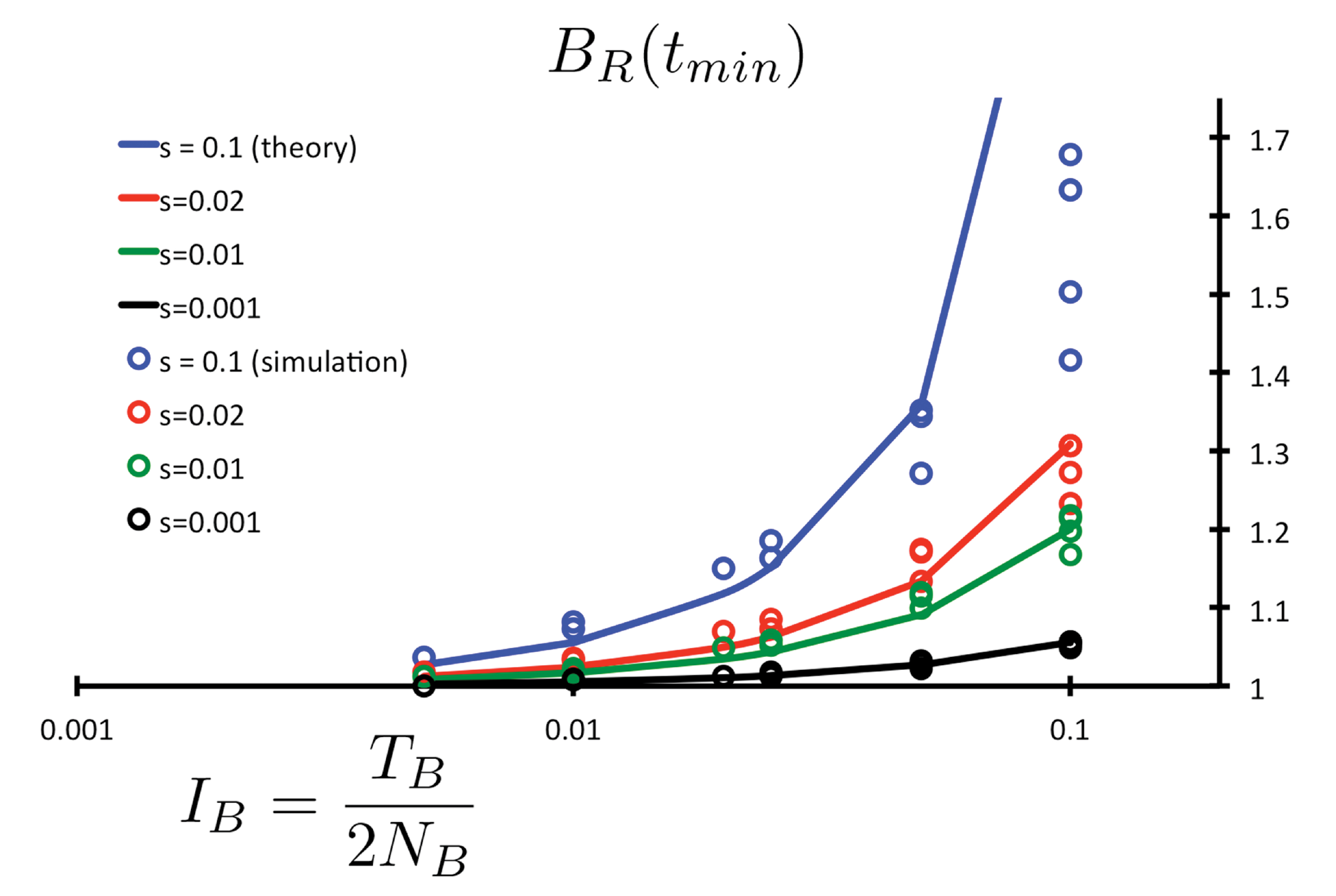}
\end{center}
\caption{{ Maximum response values of the burden ratio $\br(t_{min})$ are plotted for recessive selection as a function of bottleneck intensity.  A wide range of parameter sets are plotted with all combinations of $2N_B= \{2000, 1000, 400, 200,100\}$, $s = \{ 0.1, 0.02, 0.01,0.001\}$, $T_B = \{ 200,100, 50, 20, 10\}$  }, each simulated for $10^8$ nucleotide sites.  For relatively low intensity bottlenecks we note excellent agreement over the parameter ranges plotted.  Intensities with $I_B  = T_B/2N_B > 0.1$ are excluded, as the instantaneous bottleneck scaling breaks down in favor of a long bottleneck scaling.   The approximation necessarily weakens for simulations that represent longer bottlenecks, and only for strong selective coefficients, as expected.   This quantifies the limitations of the instantaneous bottleneck approximation, as we observe substantial deviation only around $I_B = 0.1$ and with selection strength $s=0.1$.        
\label{fig:big_collapse}}
\end{figure}

We checked our analytic results using a forward time population simulator, described in detail in the {\bf SI}.  Given the ubiquity and analytic simplicity of the exponential decay in the additive scenario, we focus here on our predictions for recessive variation.  We compare analytic expressions of $\br(t_{min})$ at the peak response given in Equation (\ref{IBmax}) for various selection coefficients.   We simulated a wide range of bottleneck parameters to test the limitations of our theoretical understanding.  In Figure \ref{fig:big_collapse}, we demonstrate the accuracy of our analytic results, by plotting the ratio of the simulated values of $\br(t_{max}, s, I_B)$ to our analytic predictions $\br(t_{max}, s, I_B)$ as presented in Equation (\ref{IBmax}).   We arrange our simulated data by bottleneck intensity $I_B$, as we expect the instantaneous bottleneck approximation to break down as intensity is increased due to longer bottleneck duration $T_B \gg 1$.  As plotted, complete agreement between simulated data and analytic predictions is represented by a flat line at $\br^{sim}/\br^{analytic} = 1$.   As expected, we find deviations as we approach the limitations of our perturbative approximation roughly around $T_b \sim 2N_B/10$ when $I_B \sim 0.1$.  Below these higher intensities, we find quite good agreement for all parameter sets well below 10\%  error, even at $I_B = 0.05$.

%
%
%
%

\section{Discussion}

The increase in prevalence of recessive phenotypes following population bottlenecks has been attracting the interest of geneticists for a long time \cite{Robertson1952,Slatkin2004}. Theoretical analysis of allele frequency dynamics in a population expanding after a bottleneck suggested that frequency of an individual allele may rise due to increased drift \cite{Slatkin2004, Gazave2013,Peischl2013}. Here, we focus on a more general question of the collective dynamics of recessively acting genetic variation. Surprisingly, our analysis suggests that the number of recessively acting variants per haploid genome is reduced in response to a bottleneck and subsequent re-expansion. Generally, we have demonstrated that the frequency spectra of recessive deleterious polymorphisms behave distinctly from additively acting variation following a population bottleneck and subsequent re-expansion. The response of additive variation depends crucially on the average number of deleterious alleles, and on the number of generations for which selection is relaxed during the bottleneck. In contrast, the dynamics of recessive variation crucially depend on the width of the site frequency spectrum, rather than the average number of mutations per individual, such that the accumulation of deleterious mutations can respond strongly even to a single generation bottleneck. Importantly, the temporal dynamics of the accumulation of deleterious alleles depends qualitatively on dominance coefficient and quantitatively on selection coefficient. The qualitative dependence on dominance coefficient allows for a robust statistical test for recessivity. If the variation is additive, the number of deleterious variants per a haploid genome is larger in a bottlenecked population than in a corresponding equilibrium population. If the variation acts recessively, this number is smaller.  The selection coefficient determines the timing of response to a bottleneck.

By explicitly analyzing the non-equilibrium response to a bottleneck, we have demonstrated a technique for using potentially confounding demographic features to probe the underlying population genetic forces.
In realistic populations, for example in modern humans, substantial work has been done to identify and understand the recent demographic history of geographically disparate populations 
\cite{Keinan2007, Lohmueller2008, Gravel2011, Fu2013, Tennessen2012, Gronau2011, Li2012, Sheehan2013, Harris2013, Macleod2013}. 
In the case of the ``Out of Africa" event, a historically substantiated and believable demographic model can be used to model the difference between African and European populations since their divergence. Comparison between populations that have and have not undergone a bottleneck can be used to infer plausible selection and dominance coefficients. In an accompanying paper \cite{Do2014}, we specialize this analysis using a realistic demographic model to attempt to bound the selection and dominance coefficients in modern human populations. Parameterizing only by the duration of the bottleneck $T_B$, along with $s$ and $h$, one can show that a substantial fraction of this three dimensional space is disallowed by the observation of even a single bottleneck.

Although the net number of recessive deleterious mutations is reduced as a consequence of a founder`s event and subsequent re-expansion, the fitness of individuals carrying these alleles is not increased, but rather decreased; selection acts only at homozygous sites and the number of homozygotes is known to increase after a population bottleneck. However, the number of heterozygous deleterious sites, or the average carrier frequency for associated alleles, is suppressed, such that the mating of individuals from disparate bottlenecked populations may result in a decreased incidence of recessive phenotypes in such mixed lineages. In studies of model organisms, this may have applications when comparing laboratory populations founded from a few wild type individuals to their corresponding natural population.

In principle, the results of this study are applicable to the analysis of specific groups of genes and pathways. Sufficiently large subsets of alleles that are medically relevant may be analyzed in humans to identify the mode of selection for candidate variants of recessive diseases. For model organisms with a significant density of deleterious alleles, it may be possible to create a Òdominance mapÓ of the genome.

In sum, the non-equilibrium dynamics induced by demographic events is an essential, and indeed insightful, feature of most realistic populations. Population bottlenecks, abundant in laboratory populations and in natural species, have the potential to provide a novel perspective on the role of dominance in genetic variation.

\section{Acknowledgements}
The authors would like to thank Benjamin Good, Alexey Kondrashov, Nick Patterson, Jonathan Pritchard, and Guy Sella for particularly useful discussions.  DJB and SS were generously supported by NIH grants R01 MH101244 and R01 GM078598.  RD was supported by a CIHR Banting fellowship.  DR is grateful for support from NIH grant R01 GM100233.
\newline

\bibliographystyle{genetics}
\bibliography{bottle}

\newpage
\section*{Supplemental Information}

\subsection*{Appendix I: Integration of the Kolmogorov Equation}
\label{subsection:kolmo}

The Forward Kolmogorov Equation is generally used to describe the probability distribution of the trajectory of an allele in frequency space within a population \cite{Kimura1964, Ewens2004}.  If one analyzes the frequency of all polymorphic alleles in the population, or the site frequency spectrum (SFS), one can describe the collective dynamics of this distribution in a very similar form using an infinite sites model.  The Kolmogorov equation describes the time dependence of the probability density $\rho(x,t)$ of an allele's frequency $x$ at some time $t$.  In the limit of a large number of simultaneously polymorphic alleles, one can think of all points in this probability distribution as being filled by one or more alleles.  Choosing to focus on the case of purely recessive variation, one can write down the time dependence of the SFS in the following form.
\beq
\label{SFS_kolmo}
\dd_t \phi(t,x) = s \dd_x \left( x^2(1-x)\phi \right) +\frac{1}{4N} \dd_x^2(x(1-x)\phi) +2 N U_{d} \  \delta \left[x-\frac{1}{2N}\right]
\eeq
The presence of the delta function represents an influx of new mutations into the spectrum at initial frequency $1/2N$ coming from $2N$ individuals in the population, each with a mutation rate $U_{d}$ per individual per generation.  This acts like a source in the SFS at $x=1/2N$, and is a reasonable approximation in the limit of a long genome with no double mutations or back mutations.  We are interested in the time dependence of specific moments of this distribution.  For example, to determine the time dependence of the first moment of the distribution $\xbar$, we multiply by $x$ and integrate to find the time dependence of this moment.
\beqa
 \dd_t \xone = \int dx \  x \ s \dd_x \left( x^2(1-x)\phi \right) + \int dx  \ x \ \frac{1}{4N} \dd_x^2(x(1-x)\phi)
 + \int dx \ x \ 2 N_0 U_{d} \  \delta \left[x-\frac{1}{2N_0}\right] \nonumber \\
\eeqa
The delta function integral is trivially computed and we integrate by parts once on each of the other integrals, noting that the boundary terms vanish due to the $x^2(1-x) \phi(x)$ factor under the derivatives, which scales as $x$ at low frequencies $x \rightarrow 0$ and decays rapidly as $x\rightarrow 1$ provided selection is efficient.
\beq
\dd_t \xone = -\int dx \ s \left( x^2(1-x)\phi \right) - \int dx  \frac{1}{4N} \dd_x(x(1-x)\phi) + U_{d}
\eeq 
The drift term can be identified as a total derivative, which vanishes, leaving the following dynamical equation for the mutation burden.  
\beqa
\dd_t \la x(t) \ra^{recessive}  = U_{d} - s \left( \langle x(t)^2 \rangle - \langle x(t)^3 \rangle \right)
\eeqa
The time dependence of all higher moments can be computed in a completely analogous way.  Since it is relevant for our present purposes, we note the equation of motion for the second non-central moment.
\beqa
\label{x2_kolmo}
\dd_t \la x^2(t) \ra^{recessive} &=& \frac{U_{d}}{2N} - 2 s \left( \langle x(t)^3 \rangle -  \langle x(t)^4 \rangle \right) + \frac{1}{2N} \left( \langle x(t) \rangle - \langle x^2(t) \rangle  \right) 
\eeqa

Equations of motion for moments of the site frequency spectrum of alleles under purely additive selection can be computed in the same way.  Here we cite these results for convenience.  Note that we are using the convention $s_{add} \deff hs = s/2$
\beqa
\label{x1_kolmo_add}
\dd_t \la x(t) \ra^{additive}  \approx U_d - \frac{s}{2}  \left( \langle x(t) \rangle - \langle x^2(t) \rangle  \right)
\eeqa
\beqa
\label{x2_kolmo_add}
\dd_t \la x^2(t) \ra^{additive} \approx \frac{U_{d}}{2N}  -  s \left( \langle x^2(t) \rangle - \langle x^3(t) \rangle  \right) + \frac{1}{2N}  \left( \langle x(t) \rangle - \langle x^2(t) \rangle  \right)
\eeqa

In the limit $ \sqrt{2N s} \gg 1$, the SFS for recessive alleles rapidly vanishes at high frequencies such that we can drop the $(1-x)$ dependence to find the following approximate equation of motion.
\beqa
\label{x1_kolmo}
 \dd_t \la x(t) \ra^{recessive}_{\sqrt{2Ns} \gg 1} \approx U_{d} - s \langle x(t)^2 \rangle \quad  \text{   for $\sqrt{2N s} \gg 1$}
\eeqa
The $(1-x)$ contribution in the dynamics of higher moments can be similarly neglected.  For alleles under additive selection, the analogous strong selection limit is given by$2N s \gg 1$, which results in the following simplified dynamics.
\beqa
\label{x1_kolmo_add}
\dd_t \la x(t) \ra^{additive}_{2Ns \gg 1}  \approx U_d - \frac{s}{2}   \langle x(t) \rangle
\eeqa
Notably, this equation of motion is diagonal and can be easily solved analytically, as is the case for all higher moments of the SFS for alleles under strong additive selection.

\subsection*{Appendix II: Analytic calculation of the trajectory of the mutation burden for recessive selection}
\label{subsection:big_calculation}

Here we are interested in the motion of the first moment $\xbart$ of the distribution $\phi(t)$ after re-expansion from the bottleneck.  First, we consider the equation of motion given by Equation (\ref{mean_motion}), which is derived above.  We repeat it here for the convenience of the reader.

\beq
\label{dtx}
\dd_t \la x (t) \ra \approx U_{d} - s \langle x(t)^2 \rangle
\eeq
Since the time scale on which the mutation burden rises to a maximum is shorter than the time scale of drift, we can imagine rescaling time by the effective population size $2N_0$ and then working in the perturbative regime $t \ll 1$.  This allows us to Taylor expand near $t = 0$ to understand the motion of the burden at early times immediately after the bottleneck.  We later determine all of the moments used below and see sufficient subsequent suppression to validate this expansion.
\beq
\label{taylor2}
\dd_t \la x (t) \ra \goes U_{d} - s \left[ \langle x(0)^2 \rangle + t \dd_t \langle x(t)^2 \rangle |_{t=0} + \frac{t^2}{2} \dd_t^2 \langle x(t)^2 \rangle  |_{t=0} + \Ocal(t^3) \right]
\eeq
To understand the time dependence of $\la x^2 \ra$, we analyze the next moment in the same fashion as employed for the first moment, as described in the previous appendix and given in Equation (\ref{x2_kolmo}).  
\beq
\label{dtx2}
\dd_t \la x ^2 \ra \approx \frac{U_{d}}{2N_0}  - 2 s \langle x^3 \rangle + \frac{2}{4N_0} \la x \ra
\eeq
Note that these moments and all higher moments have a non-negligible contribution from the diffusion term in the forward equation.   

We model an instantaneous bottleneck as a single generation downsampling of $2N_B$ chromosomes out of the original population of $2N_0$ chromosomes.
We can approximately compute $\xtwo$, $\xthree$, and higher moments if desired, immediately after bottleneck sampling (denoted ``$after$") since we have an integral form for $\phi_B(x)$ given by appropriately scaling $k$ in terms of $x$ in Equation (\ref{phib}).  Here, $\phi_0$ represents the initial pre-bottleneck site frequency spectrum, and the $n^{th}$ moment of this distribution is represented as $\la x^n \ra_0$.
\beq
\label{binom2}
\xtwo_{{after}} = \frac{1}{(2N_B)^{2}} \sum_k k^2 \ \binom{2 N_B}{k} \int dx \   (1-x)^{2 N_B - k} (x)^k \phi_0(x)
\eeq
The exchanging the order of the integral and the sum, the sum can be computed directly as a function of $x$ corresponding the the second non-central moment of the binomial distribution.  One can compute $\xthree$ completely analogously.
\beq
\label{binom3}
\xthree_{{after}} = \frac{1}{(2N_B)^{3}} \sum_k k^3 \ \binom{2 N_B}{k} \int dx \   (1-x)^{2 N_B - k} (x)^k \phi_0(x)
\eeq
The first three non-central moments of the binomial distribution are as follows:
\beqa
 \mu'_1 &=& 2N_B x \\
 \mu'_2 &=&  2N_B x(1-x) + (2N_B )^2 x^2 \\
  \mu'_3 &=& 2N_B x(1-x )(1-2x)
 + 3 (4N_B^2 x^2 (1-x) + 8N_B^3x^3)   -16 N_B^3 x^3. 
\eeqa
In the limit $N_B \gg 1$, the second and third moments are well approximated by the following expressions.
\beqa
\mu'_2 &\approx& 2N_B x+ (2N_B )^2 x^2 \\
\mu'_3 &\approx& 2N_B x + 3 (2N_B)^2 x^2 + (2N_B)^3 x^3
\eeqa
From this we can directly compute the sum in Equations (\ref{binom2}) and (\ref{binom3}).
\beqa
\xtwo_{{after}} &=& \frac{1}{(2N_B)^{2}}  \int dx \ \mu'_2  \ \phi_0(x) \nonumber \\
 &=&  \int dx \ \left( \frac{ x }{2N_B} + x^2  \right) \phi_0(x) \nonumber \\
  &=&  \frac{\xone_0}{2N_B} +  \xtwo_0
\eeqa
For the third moment, we find the following expression.
\beqa
\xthree_{{after}} &=& \frac{1}{(2N_B)^{3}}  \int dx \ \mu'_3  \ \phi_0(x) \nonumber \\
 &=&  \int dx \ \left( \frac{x}{(2N_B)^2} + \frac{3  x^2}{2N_B} +  x^3
  \right) \phi_0(x) \nonumber \\
  &=&  \frac{\xone_0}{(2N_B)^2} + \frac{3 \xtwo_0}{2N_B} +  \xthree_0
\eeqa
The third moment is relevant in that it allows us to approximately compute the time dependence of the second moment immediately after re-expansion.
\beqa
\dd_t \xtwo_{{after}} & \approx &  \frac{U_{d}}{2N_0} - 2 s \xthree_{{after}} + \frac{\xone_{{after}}}{2N_0}  \nonumber \\
&=&  \frac{U_{d}}{2N_0} - 2 s \left( \frac{\xone_0}{(2N_B)^2} +  3 \frac{\xtwo_0}{2N_B} + \xthree_0 \right) + \frac{\xbar_0}{2N_0} \nonumber \\
\eeqa
All of the $\la x^m \ra_0$ moments can be computed from the initial distribution, determining the Taylor expanded expression in Equation (\ref{taylor2}) explicitly.  These integrals are well approximated in the limit $N_0 s \gg 1$, as described in a following appendix.    We calculate the integrals using this approximation and express the first three moments below, the first two of which were described originally in \cite{Nei1968}. 
\beqa
\label{largeNS}
\xbar_0 &\like& \frac{U_{d} \sqrt{4 N_0 }}{\sqrt s} \nonumber\\
 \xtwo_0 &\like& \frac{U_{d}}{s} \nonumber \\
\xthree_0 &\like& \frac{U_{d} }{s\sqrt{4 N_0 s}}
\eeqa
Additionally, we are working under the approximation of a relatively short bottleneck, such that $\xone_{after} \approx \xone_0 + U_d T_B \approx \xone_0$.  Corrections can easily be computed to determine the $T_B$ dependence, if desired. Plugging these in, we can gauge the order of magnitude and sign of the initial contributions to the motion of the mutation burden. 
\beqa
\label{dx2dt_dependence}
\frac{\dd_t \xtwo_{{after}}}{U_{d}} &\like&  \frac{U_{d}}{2N_0}  - 2 \left( \frac{\sqrt{4 N_0 s}}{4N_B^2} +   \frac{3}{2N_B } + \frac{1}{\sqrt{4 N_0 s}} \right) + \frac{2}{\sqrt{4N_0 s}} \nonumber \\
 &\like&    -\frac{\sqrt{4 N_0 s}}{2N_B^2} -\frac{3}{N_B} 
\eeqa
Note that the $N_0^{-\h}$ terms exactly cancel in the previous equation and that we have suppressed $\Ocal(N_0^{-1})$ corrections.
Putting these results together, we integrate Equation (\ref{taylor2}) to find the following time dependence $\xbart$.
\beq
\xbart \goes \xbar_0+U_{d} t  - s t \xtwo |_{t=0}   -s \left( \frac{t^2}{2}\right)  \dd_t \xtwo |_{t = 0} + \Ocal(t^3)
\eeq
Here the integration constant is simply the initial first moment immediately after re-expansion (which is well approximated by $\xbar_{{after}} = \xbar_0$ in the case of a strong instantaneous bottleneck).  
We substitute our computed value from Equations (\ref{largeNS}) and (\ref{dx2dt_dependence}) in the above equation to compute the time dependence of the mutation burden $\xbart$.
\beq
\frac{\xbart}{U_{d}} \like \sqrt{\frac{4 N_0}{s}}\left( 1-  \frac{s t}{2N_B} \right) +  \frac{s t^2}{2N_B} \left( \frac{\sqrt{4 N_0 s}}{2N_B} + 3 \right)+ \Ocal (t^3)
\eeq
At this point, we generalize to non-instantaneous, but low intensity bottlenecks with the substitution $\frac{1}{2N_B} \rightarrow I_B \deff \frac{T_B}{2N_B}$.  By doing this we have matched the bottleneck intensity to that of a more extreme, but single generation bottleneck.  The time dependence of the mutation burden can be approximated as follows.
\beq
\xbart \like \xbar_0 \left( 1- s t I_B  +  s t^2 I_B \left( s I_B + 3 \sqrt{\frac{s}{4N_0}} \right) \right) 
\eeq
From this we can easily compute the time dependent form $\br(t) = \frac{\xbar_0}{\xbart}$.
\beq
\br(t) \like \left( 1- s t I_B  +  s t^2 I_B \left( s I_B + 3 \sqrt{\frac{s}{4N_0}} \right) \right)  ^{-1}
\eeq
This quadratic time dependence allows us to find extrema.  Note that the inclusion of higher order contributions allows for a more accurate temporal dependence $\xbart$, however this is somewhat unnecessary to understand the dominant behavior of the curve.
Concentrating just on this second order expansion in $t$, we find that the curve first drops from its initial value $\la x(0) \ra = \frac{U_{d} \sqrt{4 N_0}}{\sqrt s}$, quickly reaches a minimum, and is then brought back up by the the positive second order term.  The location of the minimum can be found approximately by solving the following equation.
\beq
\dd_t \xbartmin = 0 = - \ib \sqrt{ 4 N_0 s}   +   \ib {2 s t_{min}} \left( \ib \sqrt{4 N_0 s}+ 3 \right)
\eeq
\beq
t_{min} \sim \frac{\sqrt{\frac{4N_0}{s}}}{\left(2\ib \sqrt{4 N_0 s}+ 6\right)} \sim \h \left(s \ib  + 3\sqrt{\frac{s}{4N_0}} \right)^{-1} 
\eeq
As expected, the second derivative is positive at this extremum, implying a local minimum.
\beq
\dd_t^2 \xbartmin =  2 s \ib \left( \ib \sqrt{4 N_0 s} + 3 \right)> 0
\eeq
Plugging $t_{min}$ into our expression for $\xbart$, we find the approximate magnitude of the mutation burden at this minimum.
%
\beqa
\xbartmin \like \xbar_0 \left( 1-  \left(4 + \frac{12 }{\ib  \sqrt{4N_0 s}} \right)^{-1}  \right) 
\eeqa
We have factored out $\xbar_0 \like \frac{\theta_0}{\sqrt{4N_0 s}}$ here since it allows for easier calculation of $\br \deff \xbar_0 / \xbar$ below.  Thus, in the limit $N_0 s \gg 1$ employed to approximate $\xbar_0$, we find the following minimum value for the average number of recessive deleterious mutations per genome following a bottleneck.
\beqa
\xbartmin \like \theta_0 \left( \frac{1}{\sqrt{4N_0 s}}-  \frac{1}{\left(4\sqrt{4N_0 s} + 12/\ib \right)}  \right) 
\eeqa
From this expression, we can immediately calculate the peak value for the $\br$ statistic as follows.
\beqa
\br(t_{min})&\deff& \frac{\xbar_0}{\xbartmin}\nonumber \\
& \sim&  \left( 1-  \left(4 + \frac{12}{\ib \sqrt{4N_0 s}} \right)^{-1}  \right) ^{-1}\nonumber \\ \nonumber \\
& \sim& \frac{ 4 \ib \sqrt{4N_0 s} + 12 } {3\ib \sqrt{4N_0 s} + 12 }
\eeqa
We note that in the limit $N_B^2 \gg N_0 s $, which is biologically relevant for humans, these results simplify as follows.  The time dependence of the mutation burden for the founded population is given by, 
\beq
\frac{\xbar_{after}}{U_{d}} \sim \frac{\sqrt{4 N_0}}{\sqrt s}\left( 1- s t \ib \right) +  3s t^2\ib.
\eeq
This can be used to obtain the functional dependence of $\br(s,t)$.
\beq
\br(t) \sim \left( 1- s t \ib +  s t^2 \ib 3 \sqrt{\frac{s}{4N_0}} \right) ^{-1}\\
\eeq
In the limit $N_B^2 \gg N_0 s $, the peak response $\br(t_{min})$ occurs at a time, 
\beq
t_{min} \sim \frac{1}{6}\sqrt{\frac{4N_0}{s}}, 
\eeq
and takes the following approximate functional form.
\beq
\br(t_{min}) \sim \left( 1- \frac{ I_B \sqrt{4N_0s} }{12}  \right) ^{-1}
\eeq
We use this expression to compare to simulations in this regime of interest, with the understanding that it breaks down at relatively large bottleneck intensities.

\subsection*{Appendix III: Distribution of selective effects}

For a distribution of $s$ effects, the $s$ of maximum effect on $\br$ is dependent on the time since the bottleneck as given in Equation (\ref{smax}).  This describes the transient shift of the elevated load ratio towards smaller $s$ values.  To determine the total $\br^{observed}$ at the time of observation $t_{obs}$, one must integrate over all $s$ values present in the population.  This assumes that distinct $s$ classes for recessively acting deleterious alleles can be thought to behave independently in a well mixed, freely recombining diploid population. The distribution of selective effects for de novo mutations, $\rho(s)$, provides the appropriate weight associated with each class of selective effects, as the mutation rate for mutations of selective effect s is given by $U_d \rho(s)$.  Assuming a static distribution of selective effects, we can calculate the observed load ratio at $t_{obs}$.  For a given population, the observed mutation burden $\xbar^{obs} $ at the time of observation is the mutation burden for each class of selected effects averaged over their representative fraction of new mutations into the population.  
\beq
\xbart^{obs} = \int ds \ \rho(s)  \ \la x (s,t) \ra 
\eeq
This is true for both the equilibrium and founded populations, allowing us to compute the observed burden ratio $\br^{obs}$ as follows.
\beq
\br^{obs}(t_{obs}) = \frac{ \xbar^{obs}_{eq}  }{\xbar^{obs}_{founded}  } = \frac{ \int ds \ \rho(s)  \  \la x(s,t_{obs}) \ra _{eq} }{ \int ds \ \rho(s)   \   \la x(s,t_{obs}) \ra_{founded} }   
\eeq

The largest contribution to the load ratio at time $t_{obs}$ occurs at some effective $s_{max}$, denoted by $s^{observed}_{max}$.  The distinction here is that, although $s_{max}$ may have the largest mutation burden, it may occupy only a small fraction of the mutations present in the population when weighted by $\rho(s)$ and thus have a reduced effect on the observed burden ratio $\br^{obs}$.  The maximum contribution to the mutation burden $s^{obs}_{max}$ satisfies the following constraint.
\beq
 \dd_s\left( \rho(s) \la x(s,t) \ra_{founded}  \right)|_{s^{obs}_{max}} = 0
\eeq 
Although we remain agnostic to the distribution of selective effects in the present work, we mention that the model of an exponentially decaying distribution $\rho(s) \like e^{-\gamma s}$ is somewhat popular in the literature for theoretical, experimental, and aesthetic reasons.  As a result, the introduction of such a distribution (or more generally any monotonically decaying distribution) would produce an $s^{max}_{obs}$ value in the range, 
\beq
s_{max} > s_{obs} \ge 1/{2N_0}.
\eeq
The selective effect for which the observed change to the load ratio $\br(t_{obs})$ is maximized has suppressed signal relative to slightly lower $s$ values.  This is due to the effective rarity of high $s$ mutations in the population, as they both are introduced at a lower rate $\rho(s_{large}) < \rho(s_{small})$ and are being more efficiently purged from the population due to selection.  This indicates that the elevated load ratio $\br(t_{obs})$ may be most readily observed in the data by looking at mutations with low to intermediate selective effects, rather than those with highest effect.  Additionally, we note that the corrected equilibration time for the distribution of effects is given by the time constant associated with $s_{obs}$.

Most generally, the mutation burden will be comprised of a combination of alleles with varied selective effects and dominance coefficients.  Treatment of alleles with intermediate dominance coefficients is discussed below.  We can generalize our observed burden ratio as follows.

\beq
\br^{obs}(t_{obs}) = \frac{ \xbar^{obs}_{eq}  }{\xbar^{obs}_{founded}  } = \frac{ \int dh  \int ds \ \rho(s,h)  \  \la x(s,h,t_{obs}) \ra _{eq} }{ \int dh \int ds \ \rho(s,h)   \   \la x(s,h,t_{obs}) \ra_{founded} }   
\eeq

\subsection*{Appendix IV: Exponential expansion and more general geometries}

Exponential expansion is a general feature of many natural populations, particularly after a founding event, motivating the generalization of our analysis to such cases.  In this work, we describe the transient behavior of $\br(s,t)$, and the $s$ values that are favored as time progresses.  As a result, this behavior is extremely sensitive to exponential expansion, for example, as opposed to the simple square bottleneck model described above.  In the most general case, we may have a general time dependence for the population size after the bottleneck, which sensitively effects the $s$ values for which the burden ratio $\br$ is largest.  For an explanatory example, we will model the immediate exponential inflation of the size of both the founded and equilibrium populations after re-expansion from the bottleneck.

\beq
N_f(t) \like N_0 e^{t/a}
\eeq
We rescale time by the population size $t^I\deff t\frac{2N_0  }{2N_f} = t e^{-t/a} $, yielding exponentially slowed ``inflated" time in the decelerated frame of the fixed population size.  In this rescaled frame we can analyze the shift of the transient peak of the load ratio (in inflated time) $\br ^I(s ^I,t ^I)$ by plugging our new scaled time into Equation (\ref{smax}).

\beq
s ^I_{max}  \like \frac{2 N_0 e^{2t/a}}{10 t^2}
\eeq  
Note that this factor of $N_0$ refers to the initial population size prior to the bottleneck, and does not get rescaled due to the inflating population size.  Taylor expansion of the exponential demonstrates that there is a perturbative crossover at time $t \like 2a $.

\beq
s ^I_{max}  \propto \left( \frac{1}{t^{2}} +\frac{2}{at} + \frac{2^2}{2a^2} + \frac{2^3t}{6a^3} + ...  \right)
\eeq  
When $t \like a$, the third term in the expansion, initially the quadratic term of the exponential, finally begins to dominate over the second term in the expansion.  At this point positive powers become technically relevant in the perturbative expansion.  This is the transition between the initial transient decrease in $s_{max}$ and the exponential freezing out of the rapidly decaying large $s$ components of $\br$.    At this time, the maximum of the load ratio is given by, 

\beq
s_{max} \like \frac{2N_0 e^{2}}{10 a^2} \approx \frac{2 N_0}{a^2}.
\eeq
For very rapid inflation, $a$ is small, indicating that the dominant modes in $\br$ still exist at high $s$ values, such that $s_{max} \gg 1$.  For large $a \gg 1$, corresponding to slow, even adiabatic, expansion, the transient rapidly decays towards smaller $s$ values, such that $s_{max} \ll 1$.   Intermediate values are particularly interesting, as the rate of expansion can actually compete dynamically with the transient decay.  In this case, any intermediate selection effect may be frozen in, dominating the signature in the burden ratio $\br$.

\subsection*{Appendix V: The $\bd$ statistic and linearity}

Here we introduce a completely analogous statistic to $\br$.  An equally valid comparison between the two populations of interest, $\bd$ measures the difference in the mutation burden, rather than the ratio.
\beq
\bd \deff \left( \xone_{equilibrium} - \xone_{founded}  \right)= \left\{
        \begin{array}{ll}
            > 0 & \text{for additive selection} \\ \\
             < 0  & \text{for recessive selection}
        \end{array}
    \right.
\eeq
There are two primary distinctions between the $\bd$ and $\br$ measures.  First, the $\bd$ measure is linear and thus the magnitude provides an even comparison between additive and recessive variation.  Since it is a difference rather than a ratio, the space of allowed additive values remains in the range $\bd^{additive} \in [0,\infty]$, and recessive in the range $\bd^{recessive} \in [-\infty,0]$.  For comparison, the ratio $\br^{additive} \in [0,1]$, and $\br^{recessive} \in [1,\infty]$, which is clearly an uneven map, except in the form $\log \br$.  This linearity allows for simpler analysis of compounded statistical errors, and simpler convolution for a distribution of selective effects.
\beq
\la \bd \ra^{obs} = \int ds \ \rho(s)  \ \bd(s,t)
\eeq
One complication is that the mutation rate remains on overall scale factor, as $\bd \propto U_d$, making the magnitude of the measure quantitively more difficult to interpret in the absence of an extremely good estimate of the mutation rate.   Since the mutation rate cancels in the $\br$ statistic, the magnitude allows for qualitative inference of the dominance and selective coefficients unhindered by the imprecision of mutation rate estimates.  The time dependence of the statistic is given simply as 
\beq
\bd \sim  t \  U_d \ib \sqrt{4 N_0 s}  +  t^2 \ \ib s U_d  \left( \frac{\sqrt{4 N_0 s}}{2N_B} + 3 \right)+ \Ocal (t^3).
\eeq
In the limit of interest $\ib^{-2} \gg N_0 s$, this expressions simplifies to the following.
\beq
\bd \sim  t \  U_d \ib \sqrt{4 N_0 s}  +  t^2 \ 3 U_d s \ib + \Ocal (t^3)
\eeq

\subsection*{Appendix VI: General dominance coefficient}
\label{subsection:generalh}

The analysis above presumes that deleterious mutations act with a single average selective effect, either purely additively or purely recessively.  One can extend our analysis to the case of partial dominance with a general coefficient $1/2 \ge h \ge 0$, with extreme values corresponding to additivity and recessivity, respectively.  We ask at what value of $h$ does the crossover from $\br<1$ to $\br>1$ occur.   This crossover at some intermediate value $h=h_c$ is of practical interest, as our statistic only has sensitivity to detect whether the average dominance coefficient of a set of alleles lies above or below this critical value.   The Kolmogorov equation is easily generalized to include a general dominance coefficient.
 \beqa
 \dd_t \phi(x,t) &=& 2 N U_{d} \  \delta \left[x-\frac{1}{2N}\right]  +\frac{1}{4N} \dd_x^2(x(1-x)\phi)  \nonumber \\
&&  \ \ \ +  \ s h \dd_x (x(1-x)\phi(x,t)) +  s(1- 2h) \dd_x (x^2(1-x)\phi(x,t))
 \eeqa
 As detailed in the appendix above, we can use this equation to describe the dynamics of the mutation burden.  
\beqa
\dd_t \xone \approx  U_{d}- s_A \xbar - s_R \xtwo 
\eeqa
Here we have defined $s_A \deff sh$ and $s_R \deff s(1-2h)$ for convenience, and taken the strong selection limit in the initial and final population, such that both $2N_0 s_A \gg1$ and $\sqrt{2N_0 s_R} \gg1$ are satisfied.  In this limit, one can compute the dynamics of the moments after a short bottleneck with completely relaxed selection in complete analogy to the recessive case described above.  The perturbative dynamics immediately after re-expansion from the bottleneck are well described by the following Taylor expansion.
\beqa
\la x (t) \ra &\approx& \la x(0) \ra + t \ \dd_t \xbart |_{t=0}+ \frac{t^2}{2} \  \dd^2_t \xbart |_{t=0} + \Ocal(t^3) \nonumber \\
 &\approx& \la x(0) \ra + t  \left( U_{d} - s_A \xone -s_R \xtwo \right)|_{t=0} - \frac{t^2}{2} \left(s_A \dd_t \xone + s_R \dd_t \xtwo    \right)|_{t=0} + \Ocal(t^3)
\eeqa
The crossover value occurs when $\br = 1$, such that $\xbart_{founded} = \xone_0$, providing the following time dependent condition.  
\beq
(\la x (0) \ra -\xbar_0) + t  \left( U_{d} - s_A \xone -s_R \xtwo \right)|_{t=0} - \frac{t^2}{2} \left(s_A \dd_t \xone + s_R \dd_t \xtwo    \right)|_{t=0} \approx 0
\eeq
As described above, this can be expressed in terms of the moments of the initial distribution $\la x^n \ra_0$.  The values of $s_A$, $s_R$, and all of the moments of the initial distribution are a function of the dominance coefficient $h$, such that the solution to the above equation provides the crossover value $h_c$.  Given the exponential dependence of the the initial distribution $\phi_0(x)$ on $h$, this equation is generally transcendental and thus requires a numerical solution.  

\begin{figure}
\begin{center}
\includegraphics[width=0.9\columnwidth]{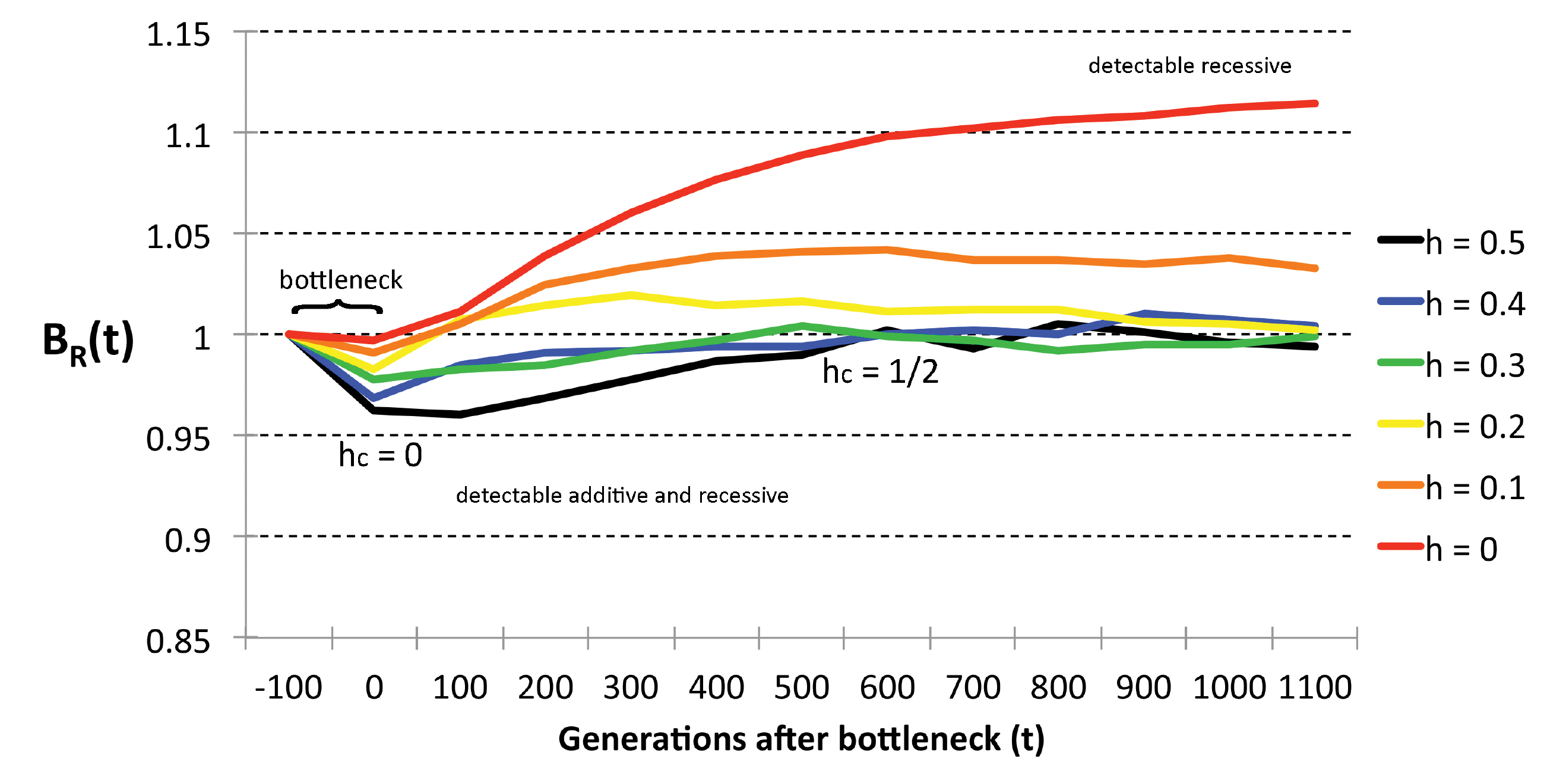}
\includegraphics[width=0.9\columnwidth]{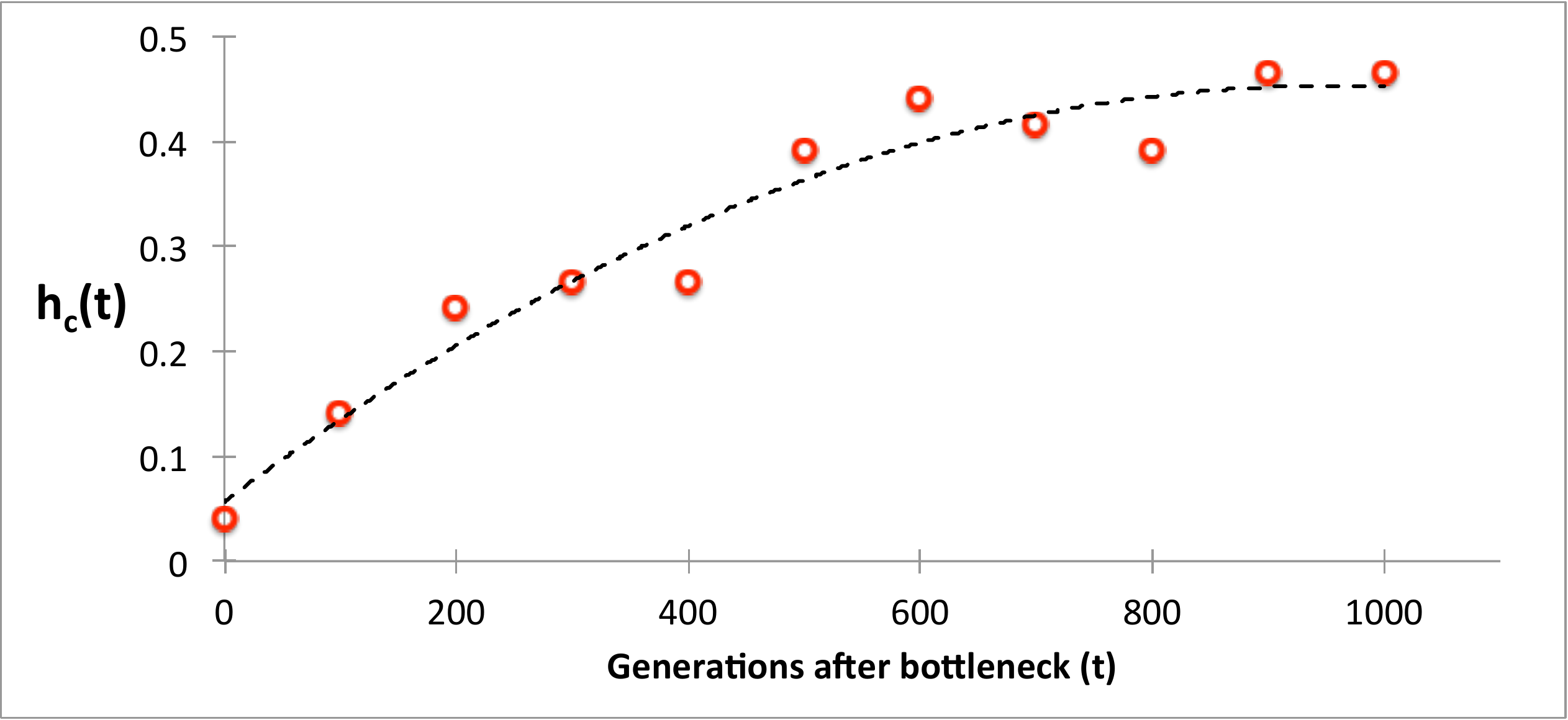}
\end{center}
\caption{{ {\bf ABOVE:} Several values of dominance coefficient $h$ are plotted as a function of time after re-expansion from the bottleneck.  Additive and recessive alleles are distinguishable when observing at early times prior to re-equilibration due to additive selection.  During the equilibration process, the critical value of the dominance coefficient $h_c(t)$ at which $\br=1$ shifts from near pure recessivity ($h_c \sim 0$) at early times to near additivity at late times ($h_c = 1/2$).  After additive re-equilibration, partially recessive alleles are still detectable ($\br > 1$) with purely recessive alleles providing the largest signature prior to their eventual equilibration.  In this plot $2N_0 = 20000$, $s = 10^{-2}$, $T_B = 100$ and $2N_B = 2000$ such that $I_B = 0.05$.  This qualitative behavior is generic for most parameter values in the short, low intensity bottleneck limit $I_B \ll 1$, however the time dependence of $h_c$ depends sensitively on these parameters.
{\bf BELOW:} The crossover dominance coefficient $h_c$ is plotted as a function of time.  At early times $h_c \sim 0$ is close to pure recessivity.  After re-equilibration of additive alleles, $h_c \sim 1/2$, such that only partially recessive alleles provide a signature.   Any value $\br > 1$ provides evidence of alleles under partially recessive selection, with the largest contribution coming from purely recessive alleles.
 }\label{fig:time_hc}}
\end{figure}

Notably, the solution $h_c(t)$ is an inherently time dependent quantity.  The additive response is largely due to accumulation of mutations due to relaxed selection during the bottleneck with subsequent decay after re-expansion.  In contrast, the recessive response occurs largely after re-expansion due to the purging of newly formed deleterious homozygotes.  As a result, at very early times the crossover occurs close to pure recessivity such that $h_c(t\rightarrow 0) \sim 0$, since $\br < 1$ for even partially additive alleles at this time.  The purely additive case equilibrates far more quickly  than the recessive case ($t^A_{relax} \propto s^{-1}$ and $t^R_{relax} \propto s^{-1/2}$), such that purely additive alleles become distinguishable from all other cases with even minor excess selection on homozygotes at late times.  After this time, nearly additive modes begin to decay, such that there is a breakdown in the definition of $h_c$ since multiple values satisfy the constraint $\xbart_{founded} = \xbar_0$.  After additive alleles have re-equilibrated, partially recessive alleles remain detectable in times $t^R_{relax} > t > t^A_{relax}$, with the strongest signal coming from purely recessive alleles at $t \ge t_{min}  \propto \sqrt{4N_0/s} $.  This behavior is summarized in Figure (\ref{fig:time_hc}).

\subsection*{Appendix VII: Long bottleneck limit}

In the case of a long bottleneck of duration $T_B \sim \Ocal(2N_B)$ generations, the bottlenecked population has had sufficient time to equilibrate into mutation-selection-drift balance with the new population size $2N_B$.  The site frequency spectrum can be written in the same form given by Equation (\ref{kimura}).  In the case of recessive variation, we find the following form during the bottleneck.
\beq
\phi(x) =  \theta_B \frac{e^{-2 N_B s x^2}}{x \left(1-x \right)}   \left[ 1 -   \frac{  \int^x_0 e^{2 N_B s x^2} }{  \int^1_0 e^{2 N_B s x^2} }\right]
\eeq
Here we have defined $\theta_B \deff 4 N_B U_{d}$.  In the limit $N_b s \gg 1$, this can be written approximately as follows.
\beq
\phi(x) \approx  \theta_B \frac{e^{-2 N_B s x^2}}{x }
\eeq
Immediately after re-expansion from the bottleneck, the first three moments of this distribution can be easily calculated using the Gaussian integrals described in an appendix below.  These can be substituted into the Taylor expanded time dependent form for $\dd_t \xbart $ in Equation (\ref{taylor}) to analyze the dynamics and solve for the functional dependence of the $\br$ statistic.   

For analysis of bottlenecks of intermediate length, a full non equilibrium description is required, but this can be well approximated by analytically patching the solutions given by the instantaneous and long bottleneck limits.  


\subsection*{Appendix VIII: Simulations and curve collapse details}
\label{section:simSI}

We performed the following analysis using a forward time population simulator, custom written in {\bf C}.  For computational speed, the simulator only keeps track of allele frequencies in a freely recombining diploid system, rather than containing full genome information.  We use an infinite sites model with a mutation rate appropriate for $10^8$ bases that represents the roughly the $30Mb$ length of the human coding genome.   Allele counts in the current generation are sampled based on the frequencies in the previous generation $p_{old}$, the selection coefficient $s$, and the dominance coefficient $h$.  We calculate the expected frequency $p_{current}$ in the current generation as: 
\beq
p_{current} = \frac{(p_{old}^2 (1+s)+p_{old} (1-p_{old} )(1+s)h)}{(p_{old}^2 (1+s)+2p_{old} (1-p_{old} )(1+s)h+(1-p_{old} )^2 )}.
\eeq
The simulator has arguments for mutation rate, $U_{d}$, adding new mutations at a probability of $U_{d}$ per base pair per generation, selection coefficient $s$, dominance coefficient $h$, a burn-in of 300,000 generations where sampling occurs every 100 generations in sped-up mode, a transition to sampling every 1 generation at 1000 generations before time $t=0$, and number of replicates.  

The code was designed to allow for flexible demographic histories, in order to accurately represent events such as the "Out of Africa" migratory event in human population genetic history.  For the purposes of comparison to our analytic results, we ran simulations for a simple, square bottleneck of varying population sizes for both the equilibrium population with size $2 N_0=2 \times10^4$ and bottlenecked populations with temporarily reduced sizes of $2N_B= \{2000, 1000, 400, 200,100\}$ for a duration of $T_B= \{ 200,100, 50, 20, 10\}$ generations.  These simulations were performed under both purely additive ($h=0.5$) and purely recessive ($h=0$) selection, for a wide range of selection coefficients including $s=\{1, 0.1, 0.02, 0.01, 0.001\}$.

Here we extend our analysis of the accuracy of our analytic results by continuing to scrutinize the comparison with our simulation.  The following should be thought of an extension to the analysis described in in the main text.

\begin{figure}[H]
\begin{center}
\includegraphics[width=0.95\columnwidth]{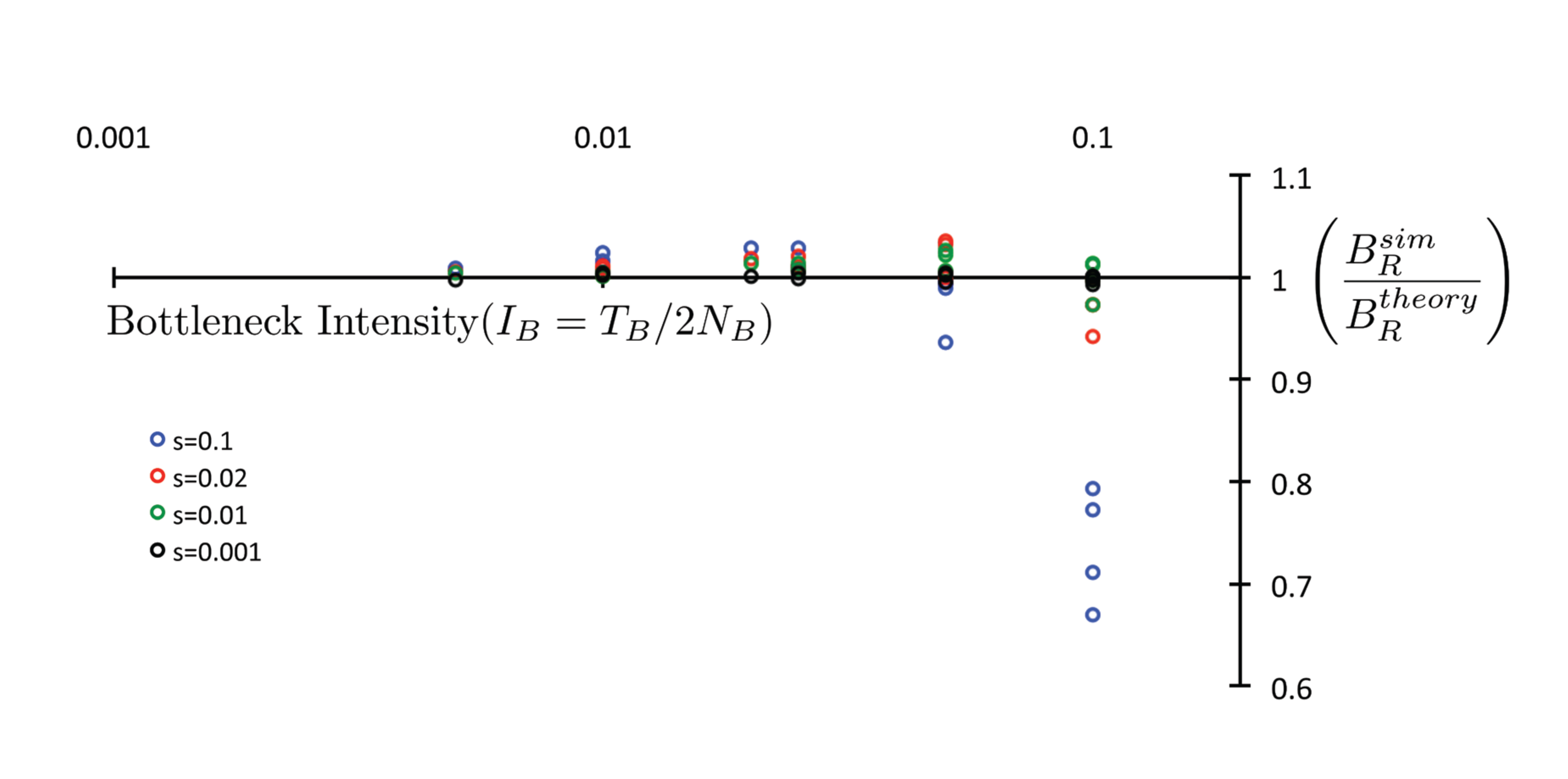}
\includegraphics[width=0.95\columnwidth]{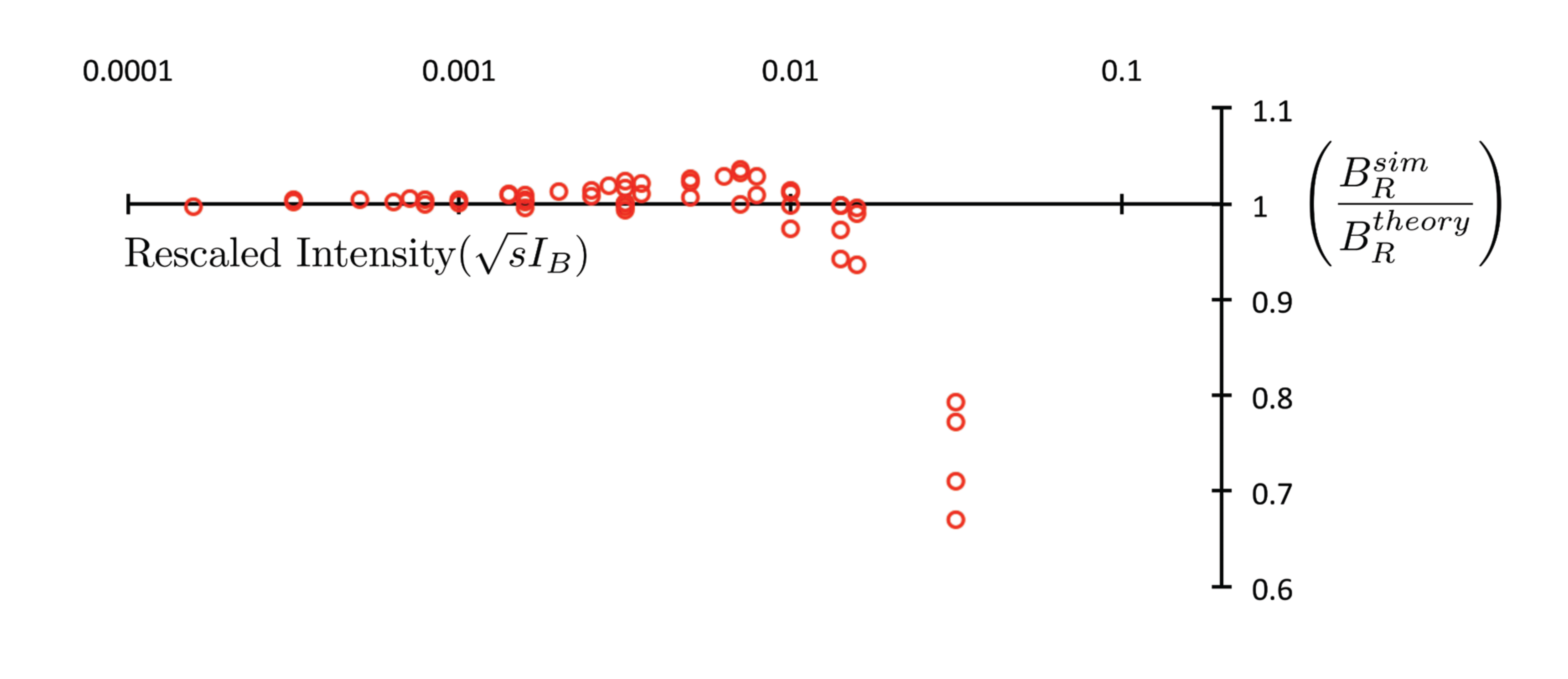}
\end{center}
\caption{{Here we plot a curve collapse to compare our analytic description to simulated data.  Values near $\br^{sim}/\br^{analytic} =1$ validate our analytic description.  Deviation from this line represents a breakdown in the proposed scaling as a function of the intensity and selective effect.  We find that the collapse is weakly stratified by selective coefficient, even in the range of good agreement at low intensity.  Large selective coefficients $s = 0.1$ deviate fastest, implying a breakdown in the short bottleneck scaling of $\br(s)$.  Parameter values of $2N_B = 2000$, $T_B = \{ 200, 100, 50, 20 \}$, and $s = \{ 0.1, 0.02, 0.01, 0.001\}$ are included on the plot.    }\label{fig:s_collapse}}
\end{figure}

To represent how the breakdown of our approximation depends on the selective coefficient, we plot a subset of the data labeled by selective effect size $s$ in Figure \ref{fig:s_collapse}.  We note that deviations from our analytic scaling occur most substantially at large selective effect as the intensity is increased, implying that the correct scaling of a more extended bottleneck involves a correction to the $s$ dependence.

\begin{figure}[H]
\begin{center}
\includegraphics[width=0.95\columnwidth]{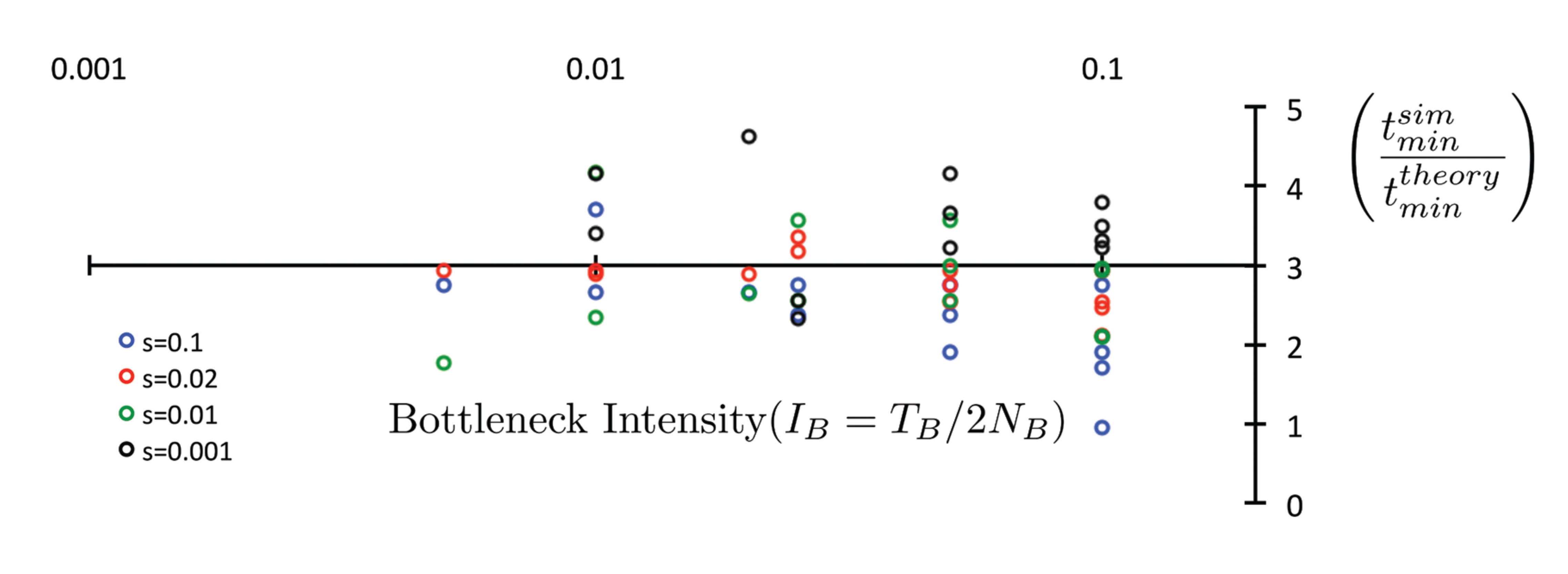}
\includegraphics[width=0.95\columnwidth]{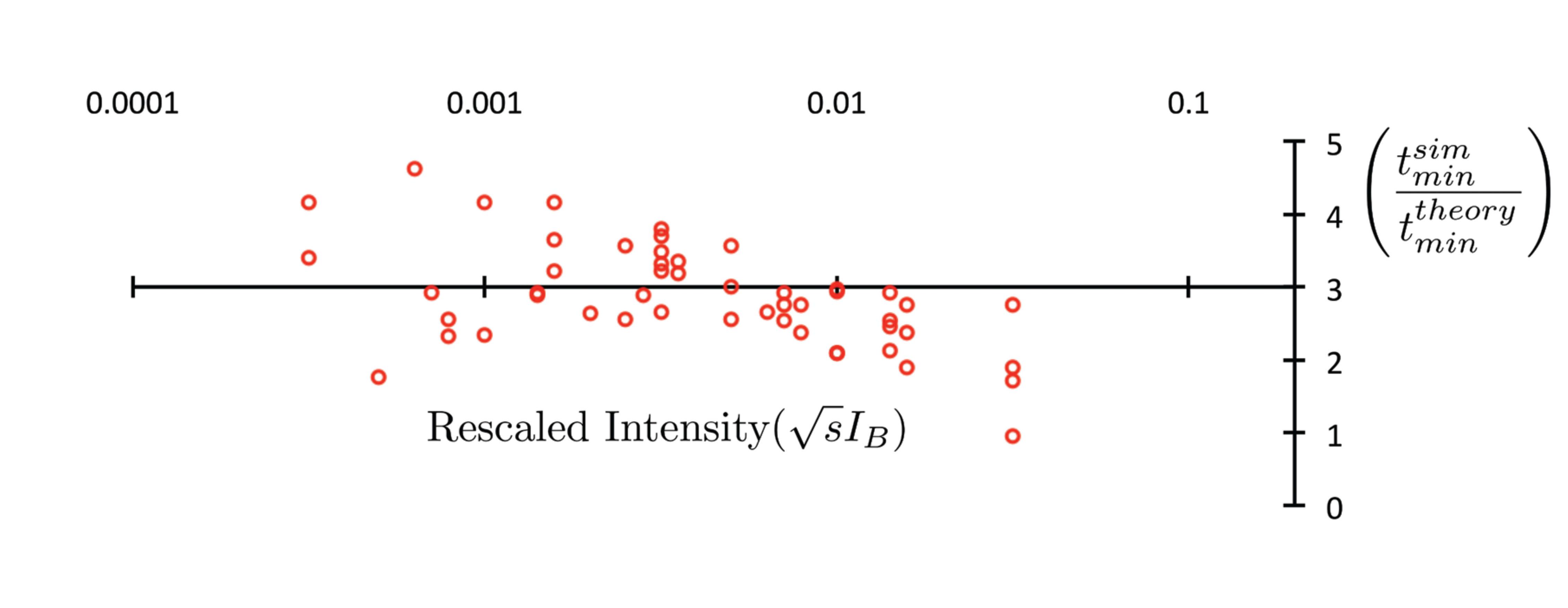}
\end{center}
\caption{{ Here we represent the collapse of the peak time $t_{min}$ by plotting $t_{min}^{sim} /t_{min}^{analytic}$.  The collapse, although very rough, appears to cluster in the low intensity regime, with relatively quick deviation as the intensity is increased.  At larger intensities, the curves appear to be approximately stratified by selective effect, with large $s = 0.1$ deviating the fastest.  The collapse occurs around $t_{min}^{sim} /t_{min}^{analytic} = 3$, implying rescaling by a constant factor.  This provides an empirical correction to the analytic dependence which has reasonable agreement with simulated results at low intensity.  We emphasize that large fluctuations in the peak time make this collapse far less informative than the $\br(t_{min})$ observable.    }\label{fig:t_collapse}}
\end{figure}

Finally, we compare the dependence of $t_{min}$ on the selective effect and bottleneck intensity in Figure \ref{fig:t_collapse}.  We find a very rough collapse at low intensity, with relatively quick deviation as the intensity is increased.  At larger intensities, the curves are again stratified by selective effect, with large $s = 0.1$ deviating the fastest.  We note here that the collapse appears to occur around $  t_{min}^{sim} = 3 t_{min}^{analytic}$, implying scaling by a constant factor of our results.  In part, this is due to various rough approximations in the integrals ($\sqrt{\pi/2}\approx 1$, for example), and can be thought of as an empirical correction to the analytic dependence that provides reasonable agreement with simulated results.   Inclusion of this factor in our analysis of $\br(t_{min})$, produces notably poorer agreement with simulation.  As is evidenced by the level of noise in Figure \ref{fig:t_collapse}, $t_{min}$ fluctuates more substantially than the magnitude of $\br(t_{min})$, making it a harder variable to use for comparison of the analytic predictions and simulated results.  In part, this is due to the coarseness of measurement only every 100 generations in our simulation.  We suggest that, when comparing to experimental sequence data, one should use the following empirically correction to the analytic dependence from Equation (\ref{tmin}) to assess the time scale of maximum response to an experimentally  observed bottleneck. 
\beq
\label{corrected_tmin}
t_{min} \like \frac{1}{2}\sqrt{\frac{4N_0}{s}}
\eeq
Importantly, this is only meant to be a rough guideline to determine the analytic parameter dependence, not an exact expression.

\subsection*{Appendix IX: Synergistic epistasis and recessive selection}

Although the statistical test described above is motivated by the desire to identify alleles under recessive selection, we note that epistatically interacting alleles may exhibit a qualitatively similar response to a population bottleneck and subsequent re-expansion.  Specifically, in the case of synergistic epistasis between deleterious variants, interacting alleles may be rapidly purged post re-expansion after rising to substantial frequency due to relaxed selection in the bottleneck.  This is precisely the case for compound heterozygotes: a population bottleneck increases carrier frequency, such that after re-expansion multiple variants may occur in the same individual.  These individuals are less fit and are subsequently purged from the population.  In this case, the selection term driving the motion of the mutation burden is proportional to $-\sum_{j} s_{ij} \la x_i x_j \ra$ for alleles with frequency $x_i$ and $x_j$ rather than proportional to $-s \la x_i^2 \ra$, however the transient dynamics are similar due to a strong quadratic dependence of the same sign.  In a sense, recessive selection can be interpreted as a subclass of alleles with synergistic effects limited to interactions between the same locus on the sister chromosome.  As a result, this statistical test should be interpreted as a way to detect non-additive negative selection, since both synergistically interacting and recessively interacting alleles provide the same qualitative signature $\br > 1$.

\subsection*{Appendix X: Evaluating Gaussian integrals}
\label{section:gaussian}

The steady state distribution prior to the bottleneck is well approximated by the following form.

\beq
\phi_0 \approx \theta_0 \frac{e^{-2N_0 s x^2}}{x(1-x)}
\eeq
The decay at large frequencies is made even more rapid by the suppressed terms, so for the present argument this form is sufficient.  Computing the first moment of this distribution corresponds to the following integral.

\beq
\xone_0  \approx \theta_0 \int_0^1 dx \  x \ \frac{e^{-2N_0 s x^2}}{x(1-x)}  \approx \theta_0 \int_0^1 dx \ \frac{e^{-2N_0 s x^2}}{1-x}
\eeq
For sufficiently large $2N_0 s\gg1 $, the exponential rapidly converges prior to reaching the $x=1$ upper limit.  In this case, in addition to canceling the linear terms in the numerator and denominator, the $(1-x)$ term in the denominator is highly suppressed by the exponential.  The first moment can be simply computed as half of a Gaussian integral.

\beq
\label{gaussian}
\xone_0 \approx \theta_0 \int_0^1 dx \ e^{-2N_0 s x^2}  \approx \frac{\theta_0}{2} \int_{-\infty}^\infty dx \ e^{-2N_0 s x^2} \approx \frac{\theta_0}{2} \sqrt{ \frac{\pi }{2 N_0 s}}
\eeq
%
%
%
%
%
Using the following definition, we can compute the first few moments of interest for the site frequency spectrum $\phi(x)$ of recessive deleterious mutations.

\beq
 \la x^{n+1} \ra_0 \propto \int_0^\infty dx \ x^n e^{-\gamma x^2 } = \left\{
        \begin{array}{ll}
             \frac{(n-1)!!}{(2\gamma)^{n/2}} \h \sqrt{\frac{\pi}{\gamma}} & \text{for even n} \\ \\
             \frac{\left(\h(n-1)\right)!}{2\gamma^{(n+1)/2}} & \text{for odd n}
        \end{array}
    \right.
\eeq

The first few moments are given by the following equations.

\beq
\xone_0 \approx \theta_0 \int_0^1 e^{-2N_0 s x^2} \goes \frac{\theta_0}{2} \sqrt{\frac{\pi}{2N_0 s}} \like \frac{\theta_0}{(4N_0 s)^{1/2}}
\eeq

\beq
\xtwo_0 \approx \theta_0 \int_0^1 x e^{-2N_0 s x^2} \goes \frac{\theta_0}{4N_0 s} 
\eeq

\beq
\xthree_0 \approx \theta_0 \int_0^1 x^2 e^{-2N_0 s x^2} \goes \frac{\theta_0}{8N_0 s} \sqrt{\frac{\pi}{2N_0 s}} \like \frac{\theta_0}{(4N_0 s)^{3/2}} 
\eeq

\beq
\xfour_0 \approx \theta_0 \int_0^1 x^3 e^{-2N_0 s x^2} \goes \frac{\theta_0}{2(2N_0 s)^2} \like \frac{2\theta_0}{(4N_0 s)^2}
\eeq

\end{document}